\documentclass[conference]{IEEEtran}
\IEEEoverridecommandlockouts
\usepackage{cite}
\usepackage{amsmath,amssymb,amsfonts,pifont}
\usepackage{algorithm}
\usepackage{algorithmic}
\usepackage{hyperref}
\usepackage{graphicx}
\usepackage[misc]{ifsym}
\usepackage{diagbox}
\usepackage{textcomp}
\usepackage[dvipsnames, svgnames, x11names]{xcolor}
\usepackage{bigstrut}
\usepackage{url}
\usepackage{booktabs}
\usepackage{multirow}
\usepackage{microtype}
\usepackage{graphicx}
\usepackage{subfigure}
\usepackage{makecell}
\usepackage{adjustbox} 
\usepackage{enumitem}
\usepackage{listings}
\usepackage{wrapfig}
\usepackage{caption}
\usepackage{bm}
\usepackage{color}
\newcommand{\ie}{\textit{i.e., }}
\newcommand{\eg}{\textit{e.g., }}
\def\BibTeX{{\rm B\kern-.05em{\sc i\kern-.025em b}\kern-.08em
    T\kern-.1667em\lower.7ex\hbox{E}\kern-.125emX}}

\makeatletter
\newcommand{\linebreakand}{%
  \end{@IEEEauthorhalign}
  \hfill\mbox{}\par
  \mbox{}\hfill\begin{@IEEEauthorhalign}
}
\makeatother
\begin{document}

\title{Scaling Law of Large Sequential \\Recommendation Models\\
\thanks{\Letter~Corresponding author.}
}

\author{
\IEEEauthorblockN{Gaowei Zhang}
\IEEEauthorblockA{\textit{Gaoling School of Artificial Intelligence} \\
\textit{Renmin University of China}\\
Beijing, China \\
wicknight2000@gmail.com}

\\
\IEEEauthorblockN{Yu Chen}
\IEEEauthorblockA{\textit{WeChat, Tencent} \\
Beijing, China \\
nealcui@tencent.com}

\and
\IEEEauthorblockN{Yupeng Hou}
\IEEEauthorblockA{\textit{University of California San Diego} \\
La Jolla, United States \\
yphou@ucsd.edu}

\\
\\
\IEEEauthorblockN{Wayne Xin Zhao~\textsuperscript{\Letter}}
\IEEEauthorblockA{\textit{Gaoling School of Artificial Intelligence} \\
\textit{Renmin University of China}\\
Beijing, China \\
batmanfly@gmail.com}

\and
\IEEEauthorblockN{Hongyu Lu}
\IEEEauthorblockA{\textit{WeChat, Tencent} \\
Guangzhou, China \\
luhy94@gmail.com}

\\
\\
\IEEEauthorblockN{Ji-Rong Wen}
\IEEEauthorblockA{\textit{Gaoling School of Artificial Intelligence} \\
\textit{Renmin University of China}\\
Beijing, China \\
jrwen@ruc.edu.cn}
}

\thispagestyle{plain}
\pagestyle{plain}
\setlength{\textfloatsep}{7pt}

\maketitle

\begin{abstract}
Scaling of neural networks has recently shown great potential to improve the model capacity in various fields. Specifically, model performance has a power-law relationship with model size or data size, which provides important guidance for the development of large-scale models. However, there is still limited understanding on the scaling effect of user behavior models in recommender systems,  where the unique data characteristics (\eg data scarcity and sparsity) pose new challenges to explore the scaling effect in recommendation tasks. 

In this work, we focus on investigating the scaling laws in large sequential recommendation models. 
Specially, we consider a pure ID-based task formulation, where the interaction history of a user is formatted as a chronological sequence of item IDs. We don't incorporate any side information (\eg item text), because we would like to explore how scaling law holds from the perspective of user behavior. With specially improved strategies, we scale up the model size to 0.8B parameters, making it feasible to explore the scaling effect in a diverse range of model sizes. As the major findings, we empirically show that scaling law still holds for these trained models, even in data-constrained scenarios. We then fit the curve for scaling law, and successfully predict the test loss of the two largest tested model scales. 
Furthermore, we examine the performance advantage of scaling effect on five challenging recommendation tasks, considering the unique issues (\eg cold start, robustness, long-term preference) in recommender systems. We find that scaling up the model size can greatly boost the performance on these challenging tasks, which again verifies the benefits of large recommendation models.  

\end{abstract}


\section{INTRODUCTION}


Scaling law, referring to the principle that describes the relationship between model performance and model/data size, has been widely studied in many fields such as natural language processing~(NLP) and computer vision~(CV). It provides an important guideline for designing and optimizing models. Recent work~\cite{openai2023gpt4,zhao2023survey,gong2023multimodal} has demonstrated that large-scale models especially transformer-based models with billions, even trillions, of parameters can achieve remarkable performance~(\eg GPT-4~\cite{openai2023gpt4}, LLaMA~\cite{touvron2023llama}). Moreover, several studies also explore scaling laws in real-world applications~\cite{henighan2020scaling,ardalani2022understanding,kaplan2020scaling,zhai2022scaling}, to yield larger performance improvement over normal-sized models and deliver better quality of service to customers. 

Inspired by these progresses, we focus on investigating how  scaling law can be employed to guide the improvement of sequential recommender systems. Specifically, we aim to explore whether larger sequential recommendation models can lead to a more significant increase in recommendation accuracy. 
In terms of data format, the task of sequential recommendation formulates interaction logs of item IDs into \emph{chronological sequences}, and the final goal is to accurately predict the IDs of the future items that a user is likely to interact with. 
Despite that previous studies have examined the scaling effect in rich-feature click-through-rate~(CTR) models~\cite{ardalani2022understanding,guo2023embedding,chitlangia2023scaling} and text-based recommendation models~\cite{shin2023scaling}, it still lacks detailed study on scaling ID-based sequential recommendation models. The comparisons of our work and other related scaling studies are presented in Table~\ref{tb:scaling-method}. 

Indeed, in NLP, language models are also trained on sequence data consisting of text tokens. Due to the common  sequential nature~\cite{hou2022towards,li2023text,hou2023learning},  it has been found that modeling \emph{behavioral sequences} in recommender systems and modeling \emph{token sequences}  in NLP are closely related.  Given the great success of scaling law in NLP, it is promising to explore the scaling law in sequential recommendation models,  and further investigate how it differs in these two different domains. 

However, to explore scaling laws of sequential recommendation models, we are faced with challenges in recommender systems, \ie the interaction data is highly sparse and noisy. 
In language modeling, it has been found that the data-constrained setting might lead to a specific scaling pattern~\cite{muennighoff2023scaling}.
Thus, it will be meaningful to investigate how the data characteristics of recommender systems affect the effect of scaling law, which  remains under-explored. Further, it is unclear whether larger models would actually outperform smaller models in recommendation tasks, especially in complex scenarios with sparse or noisy input. 

To this end, in this paper, we aim to investigate the scaling behaviors of conventional ID-based sequential recommendation~(SR) models. To align with the study in language modeling, we adopt the decoder-only transformer architecture as the backbone to explore our scaling study, which recovers the interaction sequence by predicting the next item ID conditioned on the historical interaction data. We empirically observe that scaling up models often comes with training instability. Inspired by recent work on stable training~\cite{liu2023dropout}, we develop a scalable training procedure containing two major training strategies: \textit{layer-wise adaptive dropout} and \textit{switching optimizer strategy}, so as to achieve more stable training for large-scale SR models.


By carefully setting up the experiments, we explore the scaling property of transformer models in sequential recommendation ranging from 98.3K to 0.8B parameters, and find that scaling law actually holds for the studied model scales in sequential recommendation, even in a highly data-constrained setting. 
We also conduct experiments on predictable scaling~\cite{openai2023gpt4}, which aims to predict the performance of a larger model from the performance of smaller models. Specifically, we successfully predict the performance of a 0.8B large model using several small~(\textless 100$\times$) models' performance. From the data perspective, we observe that 
large-scale models are highly data-efficient and increasing the data size is helpful to avoid overfitting. Further, we also study the effect of model shape, and show that it has a weak impact on the model performance compared to model size, indicating that extensive hyper-parameter searching may not be necessary. 

In addition to these overall findings, in recommender systems, we are more concerned with the performance advantage of large models in real-world tasks.  Since the interaction data tends to be highly sparse or noisy, it poses unique challenges for models to attain decent performance under complex recommendation scenarios.  
For this purpose, we design five challenging task settings for sequential recommendation, including long-tailed item recommendation, cold-start user recommendation, multi-domain transfer, robustness challenge, and long-term trajectory prediction. 
Our empirical experiments show that large models are consistently better than small models on all five recommendation tasks, showing a great advantage by scaling up recommendation models.  

\begin{table}[t]
\centering
\tabcolsep=0.3em
\caption{Comparison between our work and other scaling studies in recommender systems. ``ID'' denotes ID-based models, ``w/o fea'' denotes it does not rely on user profiles or item features to assist in behavioral modeling, ``Seq'' denotes sequential behaviors modeling, ``Arch'' denotes model architecture, ``DS'' denotes data scaling, ``ITT'' denotes whether it proposes improved training techniques for scaling up models, ``CRT'' denotes whether it has been tested on various complex recommendation tasks such as cold-start task.}\label{tb:scaling-method}
\begin{tabular}{@{}lccccccc@{}}
\toprule
\textbf{Method}           & \textbf{ID}     & \textbf{w/o fea}    & \textbf{Seq} & \textbf{Arch} & \textbf{DS} & \textbf{ITT} & \textbf{CRT}\ \  \\ \midrule
Guo~\emph{et al.}~\cite{guo2023embedding} & \color{Green}\checkmark  & \color{red}\ding{53}   & \color{red}\ding{53} & MLP & \color{red}\ding{53} & \color{Green}\checkmark & \color{red}\ding{53}\ \ \     \\
Ardalani~\emph{et al.}~\cite{ardalani2022understanding} & \color{Green}\checkmark  & \color{red}\ding{53}   & \color{red}\ding{53}  & MLP  & \color{Green}\checkmark  & \color{red}\ding{53} & \color{red}\ding{53}\ \ \     \\
Chitlangia~\emph{et al.}~\cite{chitlangia2023scaling}            & \color{Green}\checkmark & \color{red}\ding{53} & \color{Green}\checkmark   & Decoder   & \color{Green}\checkmark  & \color{red}\ding{53} & \color{red}\ding{53}\ \ \    \\
Shin~\emph{et al.}~\cite{shin2023scaling} & \color{red}\ding{53}  & \color{Green}\checkmark   & \color{Green}\checkmark & Encoder  & \color{red}\ding{53}  & \color{red}\ding{53} & \color{Green}\checkmark\ \ \     \\ 
Ours & \color{Green}\checkmark  & \color{Green}\checkmark   & \color{Green}\checkmark  & Decoder   & \color{Green}\checkmark  & \color{Green}\checkmark & \color{Green}\checkmark\ \ \     \\
\bottomrule
\end{tabular}
\end{table}


To summarize, the main contributions of this work are threefold: 

$\bullet$ We successfully scale decoder-only transformer-based recommendation models up to 0.8B parameters, and achieve stable performance improvement by specially designed training strategies. To our knowledge, it is the first scaling study that is built on pure ID-based recommendation models.  

$\bullet$ We conduct extensive experiments to explore the scaling effect in sequential recommendation models. We successfully fit the scaling law where test loss varies with model size in recommender systems. Moreover, we find that the scaling law holds even in data-constrained scenarios and it exhibits a weak dependence on model shape. 

$\bullet$ We further examine the scaling effect on five challenging task settings. Experiment results show that large models exhibit enhanced robustness, effectively mitigating challenges such as cold-start problems. Additionally, large models are more capable when confronted with complex tasks such as multi-domain transfer and user trajectory prediction.


\section{RELATED WORK}

In this section, we review the related work in three major aspects, namely sequential recommendation, scaling law, and large language models~(LLMs) for recommendation. 

\subsection{Sequential Recommendation} 

For the past decades, recommender systems~\cite{he2020lightgcn,kang2018self,sun2019bert4rec,hou2022towards,lin2022improving} have become crucial components of online services for routing suitable information sources to users.  As a special setting,  sequential recommendation~(SR) has attracted much attention from the research community, since users' interests are dynamically evolving over time. Early studies on SR often utilize Markov chains~(MCs) to model users' sequential behaviors~\cite{rendle2010factorizing}. With the recent advancements in deep learning, a number of deep neural network models have been developed for sequential recommendation tasks~\cite{hidasi2015session,kang2018self,sun2019bert4rec,tang2018personalized}. These models are built based on various architectures such as Convolutional Neural Networks (CNN)~\cite{tang2018personalized}, Recurrent Neural Networks (RNN)~\cite{hidasi2015session}, Graph Neural Networks~(GNN)~\cite{wu2019session,chang2021sequential} and Multilayer Perceptron~(MLP)~\cite{zhou2022filter}. Recently, full attention based transformer models~\cite{vaswani2017attention} have also been applied to SR, leading to state-of-the-art performance~\cite{kang2018self,sun2019bert4rec,li2021lightweight,hou2022core}. As a representative model, SASRec ~\cite{kang2018self} first employs a Transformer block to incorporate self-attention mechanisms. Despite that transformer models are widely explored for sequential recommendation~\cite{kang2018self,sun2019bert4rec,fan2021lighter}, most of the existing research mainly focuses on small-sized models~(for example there are typically only two layers in SASRec~\cite{kang2018self} and BERT4Rec~\cite{sun2019bert4rec}). 
It is meaningful to study how the model performance would change as the model size scales up.   

\subsection{Scaling Law} 

In various fields of artificial intelligence, scaling law has been proven universal to describe the relationship between model performance and related model factors (\eg model size, data size)~\cite{mhaskar1996neural,brown2020language,kaplan2020scaling}, especially the recent success for scaling pre-trained language models~\cite{zhao2023survey}. 
Following the scaling law, prior efforts have demonstrated the potential of scaling up both the model size and data size to achieve remarkable performance in various downstream tasks~\cite{henighan2020scaling,zhai2022scaling}. In addition, one can predict the performance of large models from the performance trends spanning by small models, so-called \emph{predictable scaling}~\cite{openai2023gpt4}. 
In particular, scaling law in recommender systems has also attracted increasing attention. Guo~\emph{et al.}~\cite{guo2023embedding} and Ardalani~\emph{et al.}~\cite{ardalani2022understanding} have explored the scaling laws in Click-Through Rate~(CTR) recommendation task. They characterize scaling efficiency for both \textit{embedding} and \textit{non-embedding} parameters and show that parameter scaling is out of steam for current CTR models. Chitlangia~\emph{et al.}~\cite{chitlangia2023scaling} model user advertisement activity sequences and inspect its scaling properties with various features. In a recent study~\cite{shin2023scaling}, Shin~\emph{et al.} scale up pre-trained SR models with textual user behavior logs and successfully improve the generalization performance. However, little work has explored whether the scaling law can still hold in conventional ID-based (the mainstream data format for developing recommender systems in the research community) sequential recommendation, where additional features are often not available in real scenarios. 

\begin{figure*}[t]
    \centering
    \includegraphics[width=1.0\textwidth]{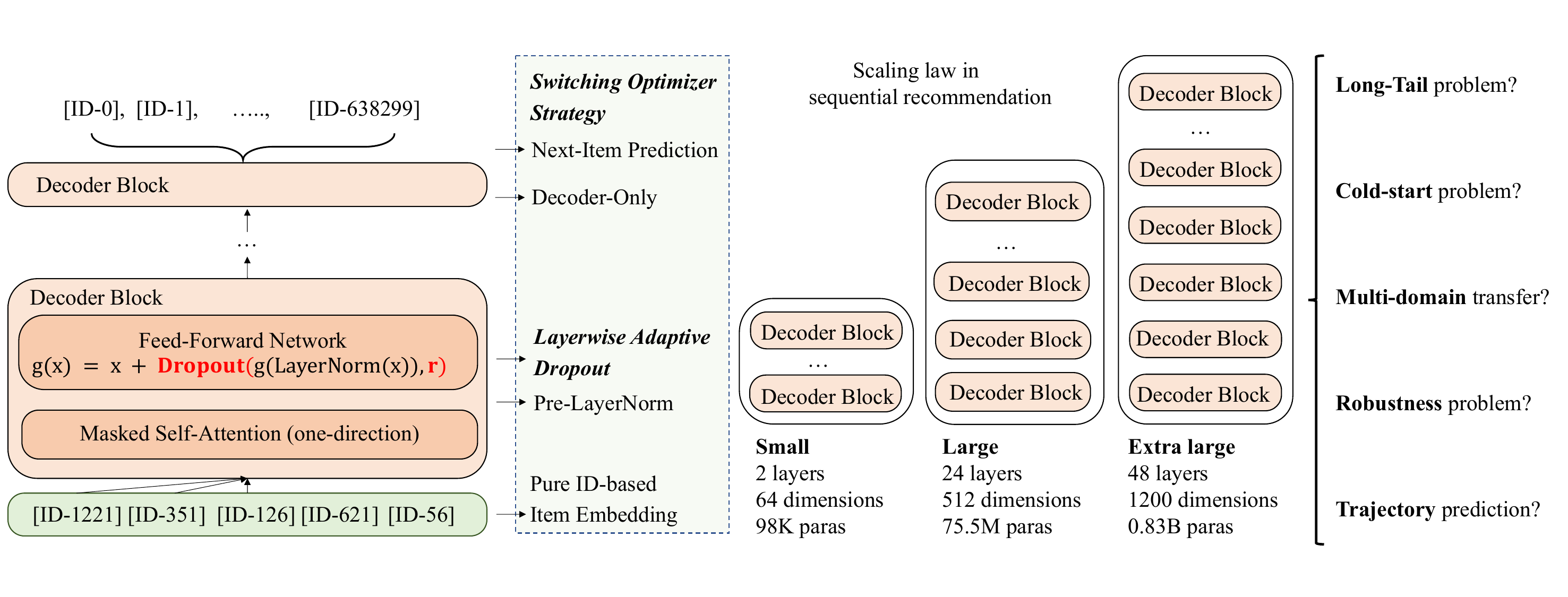}
    \caption{The illustration of the model architecture and the scaling versions.} 
    \label{fig:train}
\end{figure*}

\subsection{LLMs for Recommendation}
Recent years have witnessed the success of pre-trained language models, especially large language models~(LLMs) with an extremely large model size~\cite{zhao2023survey}. 
LLMs have exhibited an excellent model capacity and thus greatly improve the task performance in various domains~\cite{tang2023does,nov2023putting,malinka2023educational,yang2023fingpt,sun2023short,zhang2023one}. 
Especially, in recommender systems, people are trying to leverage the intrinsic world knowledge and strong reasoning capacities of LLMs to capture user preferences and deliver better ranking results~\cite{wu2023survey,lin2023can,fan2023recommender,li2023large}.
Due to the sequential nature of Transformer in LLMs, a straightforward way of using LLMs is to model sequential behaviors, \ie sequential recommendation~\cite{hou2023large,zhang2023recommendation,wang2023zero,liu2023chatgpt,harte2023leveraging,bao2023tallrec,zhang2023agentcf}.
There are mainly two paradigms to use LLMs as sequential recommendation models: (1) prompting LLMs in a zero-shot manner~\cite{hou2023large,zhang2023recommendation,wang2023zero,liu2023chatgpt} and (2) fine-tuning 
LLMs' parameters on a set of training examples to adapt LLMs on the user instructions and recommendation task~\cite{harte2023leveraging,bao2023tallrec}.
In the first paradigm, LLMs are used as ranking models in an instruction-following paradigm. User interaction history and candidate items are integrated into prompts in the form of natural language. LLMs then perform the ranking task and output results with the understanding of prompts and internal knowledge~\cite{hou2023large,zhang2023recommendation,wang2023zero,liu2023chatgpt}. In the second approach, training examples are first constructed in the form of the above prompt. LLMs are then fine-tuned on these examples and learn new domain knowledge~\cite{harte2023leveraging,bao2023tallrec}. Although some large-scale LLM-based recommendation methods are proposed, as language model backbones, LLMs' inputs and outputs are entirely text data.
Although promising, a more fundamental way in recommender systems is to index the users and items as IDs. 
There exists a natural semantic gap between text-based semantics and ID-based behavior semantics.

Our work is highly built on prior efforts on transformer-based recommendation models and scaling law for LLMs. We aim to investigate to explore how to scale up transformer models for sequential recommendation and study how scaling law applies to such models. 
We believe that this work will be useful to the design of large-scale recommendation models, and thus better model and understand user behaviors in recommender systems.

\section{Experimental Setup}
In this section, we first introduce the model architecture and datasets for exploring the scaling effect of sequential recommendation models. Further, since we empirically find that it is difficult to train large-scale transformer models for recommendation tasks, we propose improvement strategies for stabilizing the model training. 


\subsection{Model Architecture}

Following the prior empirical study in NLP~\cite{kaplan2020scaling,henighan2020scaling}, for all experiments, we adopt the decoder-only transformer models as the backbone. However, unlike language models, we don't perform tokenization on item IDs, but instead take the item set as the vocabulary for input embeddings $\mathbf{E} \in \mathbb{R}^{n \times d}$, where $n$ is the total number of items and $d$ is the latent dimensionality. Further, learnable position embeddings $\mathbf{P} \in \mathbb{R}^{s \times d}$ are also injected into input embeddings for modeling the sequential information. Here, $s$ denotes the maximum length of sequences and we set it to 50 for both datasets. Sequences exceeding this length will be truncated, and sequences shorter than this length will be padded. For a given item $v_i$, whose corresponding position is $i$, its representation can then be composed of the corresponding item embedding ${\mathbf{e}}_{v_i}$ and position embedding ${\mathbf{p}}_i$, where ${\mathbf{e}}_{v_i} \in \mathbf{E}$ and ${\mathbf{p}}_i \in \mathbf{P}$. 

After the embedding layer, we stack multiple Transformer decoder blocks. In each decoder block, multi-head self-attention is first used to aggregate items’ embeddings. At each layer $l$, query $\mathbf{Q}$, key $\mathbf{K}$ and value $\mathbf{V}$ are projected from the same input hidden representation matrix ${\mathbf{H}}^l$. The results from multiple attention heads are then concatenated and output through a learnable projection matrix $\mathbf{W^O}$:

\begin{equation}
\begin{aligned}
     {\mathrm{head}}_i  & = \mathrm{Attention}({\mathbf{Q}}_i, {\mathbf{K}}_i, {\mathbf{V}}_i) \\
     \mathrm{MultiHead}({\mathbf{H}}^l) & = [{\mathrm{head}}_1;{\mathrm{head}}_2;\ldots;{\mathrm{head}}_h]\mathbf{W^O}
\end{aligned}
\end{equation}

Due to the nature of item sequences, each input item can only attend to the past tokens and itself with the unidirectional attention mask. Then a position-wise feed-forward network with a GeLU activation is applied to increase the nonlinear capability for models:

To summarize, the overall architecture follows a typical causal decoder, the same as SASRec~\cite{kang2018self}. We implement the decoder with huggingface~\cite{wolf-etal-2020-transformers} library and conduct training and evaluation experiments on a popular open-source recommendation library RecBole~\cite{zhao2021recbole}.


\begin{table}[t]
\centering
\tabcolsep=0.8em
\caption{Statistics of the datasets after preprocessing.}\label{tb:new-dataset}
\begin{tabular}{@{}crrr@{}}
\toprule
\textbf{Dataset}           & \textbf{\#Users}     & \textbf{\#Items}    & \textbf{\#Interactions}\ \  \\ \midrule
MovieLens-20M            & 138,493 & 26,427 & 18,476,840\ \ \    \\
Amazon~(mix) & 367,710  & 240,320   & 21,787,957\ \ \     \\ \bottomrule
\end{tabular}
\end{table}

\subsection{Training Techniques}\label{subsec:train-tec}

To train our model, we optimize the model to predict the next item at time step $t+1$, conditioned on the previous $t$ items. 
Following~\cite{he2016deep,kang2018self}, we add residual
connections to propagate low-layer features to higher layers. We use pre-LayerNorm~\cite{baevski2018adaptive} for stabilizing neural network training. To alleviate overfitting issues, we also use dropout~\cite{srivastava2014dropout} for regularization. Different from the commonly used fixed dropout, we propose to use \textit{layer-wise adaptive dropout} to strike the balance between underfitting and overfitting, which will be discussed in detail in Section~\ref{subsec:train-tec}. 
For overall training setups, we use cosine learning rate schedules and weight decay for training following~\cite{hoffmann2022training,muennighoff2023scaling}. However, we find that too large weight decay will lead to model underfitting in recommender systems. Therefore, we set the weight decay to $1 \times 10^{-8}$. This is different from the settings in language models~\cite{zhao2023survey}, which we assume may be due to data sparsity. We also investigate the risk of overfitting like in~\cite{muennighoff2023scaling}, and the actual training epochs and other training hyper-parameters are detailed in Table~\ref{tb:scaling-model}. We also investigate the possible impacts of other hyper-parameters combinations, and we leave a detailed discussion of this part in Section~\ref{subsec:shape}. Overall, it is difficult to train large SR models due to the unstable training of transformer models. In our work, we propose the following improved training strategies.   

\begin{algorithm} 
	\renewcommand{\algorithmicrequire}{\textbf{Input:}}
	\renewcommand{\algorithmicensure}{\textbf{Output:}}
	\caption{Training Process of Large-Scale Sequential Recommendation Model (LSRM)} 
	\label{alg:train} 
	\begin{algorithmic}[1]
		\REQUIRE{model parameters $\theta$, layer-wise dropout rate configs $d=\{d_0, d_1,..d_n\}$ for n-layers} 
	    \ENSURE{Optimized model parameters $\theta^t$}
            \STATE{Initialize $\theta^0$, set layer-wise dropout with $d$}
		\REPEAT
                \STATE Update $\theta^0$ with Adam Optimizer
            \UNTIL Convergence~(switchover point), $\theta^c$
            \REPEAT
                \STATE Update $\theta^c$ with SGD optimizer
            \UNTIL Convergence, $\theta^t$
	\end{algorithmic} 
\end{algorithm}

\subsubsection{Layer-wise Adaptive Dropout}

When scaling up SR models, we often encounter the problem of training instability, and it is also difficult to strike a balance between underfitting and overfitting. For this issue, using different dropout rates at different training stages has been proven to improve generalization and stability~\cite{liu2023dropout}. Inspired by this finding, we adopt layer-wise adaptive dropout ratios in the overall training process. 
Specifically, we set larger dropout rates for the lower layers, as these layers directly process the primary information from data, which is a relatively simple process, and we need to prevent overfitting. Conversely, we set smaller dropout rates for the higher layers, as these layers take semantic information from lower layers and process it into more abstract representations, and we need to prevent information loss and mitigate underfitting. We conduct an ablation study on this technique and the results are shown in~\ref{tb:training-cmp}. We can find that the performance of large-scale models has declined to a certain extent without layer-wise adaptive dropout, while the performance of small models remains relatively stable. This suggests that a fixed dropout strategy is inadequate for scaling up models, making it challenging to balance between under-fitting and over-fitting and maintain stable training.

\subsubsection{Switching Optimizer Strategy}

During the training process, we experimented with a large number of learning rate schedules and optimizer choices. We empirically find that the commonly used Adam~\cite{kingma2014adam} optimizer did not perform best {in the final convergence loss}. Specifically, Adam performs well in the initial training stage, while its final convergence loss value is higher than SGD~\cite{robbins1951stochastic}. To alleviate this issue, we adopt \textit{switching optimizer strategies} following~\cite{keskar2017improving}. Specifically, we switch the optimizer from Adam to SGD at the switchover point. Prior work shows that the switchover point is also needed to learn in the training process~\cite{keskar2017improving}. However, we find that different switchover points have little impact on our datasets. Therefore, we can use the Adam optimizer convergence point as the switchover point. Based on this finding, we switch to SGD optimizer in the second training stage, without explicit learning of the switchover point. We conduct an ablation study on this technique and the results are shown in~\ref{tb:training-cmp}. We can find that the performance of models of different scales has declined to a certain extent without the switching optimizer strategy. This suggests that relying solely on the Adam optimizer to converge the model is challenging. Moreover, we also find that different switchover points have minimal impacts on recommendation performance, indicating that there is no need to explicitly identify switchover points during the training process.

\begin{table}[t]
\centering
\tabcolsep=0.6em
\renewcommand{\arraystretch}{1.2}
\caption{Ablation analysis of improved training techniques on MovieLens-20M. $\mathcal{L}_{\text{CE}}$ denotes cross-entropy loss on the test dataset. LAD denotes layer-wise adaptive dropout, SO denotes switching optimizer strategy. ${\mathrm{LSRM}}_{l_{2}}$ and ${\mathrm{LSRM}}_{l_{48}}$ denote 2-layer and 24-layer large sequential recommendation models respectively, and their corresponding hyper-parameters are shown in Table~\ref{tb:scaling-model}. All models are trained until the cross-entropy loss on the validation set no longer decreases.}\label{tb:training-cmp}
\begin{tabular}{@{}lcc@{}}
\toprule
     \diagbox[dir=SE]{Train}{$\mathcal{L}_{\text{CE}}$~$\downarrow$}{Model}   &${\mathrm{LSRM}}_{l_{2}}$   &${\mathrm{LSRM}}_{l_{24}}$   \ \  \\ \midrule
Both           & 5.6249 & \textbf{4.7182}\ \ \    \\
w/o LAD           & 5.6127 & 4.7296\ \ \    \\
w/o SO           & 5.6281 & 4.7230\ \ \    \\
None           & \textbf{5.6013} & 4.7504\ \ \    \\ \bottomrule
\end{tabular}
\end{table}

\subsection{Training Data}\label{subsec:train-data}
To train large-scale recommendation models, it is necessary to prepare sufficient sequential interaction data for learning the model parameters. 

We conduct extensive experiments on two very large public datasets in real-world recommendation scenarios, namely  MovieLens-20M~\cite{harper2015movielens}  and Amazon-2018~\cite{ni2019justifying}. 
There are a total of 29 domains in Amazon dataset. To simulate a practical recommendation scenario, for the Amazon dataset, we mix the interaction records in all domains by users and sort them by the interaction timestamps ascendingly. In this way, we can organize the interaction records into a sequential format for training transformer models. 
Although these items are attached with rich auxiliary information (\eg item title and category labels), we only preserve item IDs for sequence modeling. 
Different from language models and prior scaling law studies in recommender systems, we would like to examine the scaling law in a pure ID-based sequential setting, which has been seldom studied in the existing literature. 

To preprocess our dataset,  we follow prior studies~\cite{yu2020semi} to adopt the 30-core filtering for both users and items in the Amazon dataset, enhancing its robustness for our analysis. A difference with language models is that we don't perform tokenization (\eg BPE~\cite{gage1994new}) on item IDs, because item IDs are specific to different application platforms and it is difficult to obtain transferable units across different platforms.  

The statistics of these datasets after preprocessing are summarized in Table~\ref{tb:new-dataset}. As we can see, the entire volume of interaction records is significantly smaller than that of available text tokens in language models~\cite{zhao2023survey,brown2020language}. This is a highly data-constrained setting for sequential recommendation models.
We will further discuss the impact of data amount on the model performance in Section~\ref{subsec:data-scaling}. 

\begin{table}[t]
\centering
\tabcolsep=0.8em
\caption{Training hyperparameters of five-scale SASRec models. \#Paras~(non-emb) denotes total \textit{non-embedding} parameters, $n_{layer}$ denotes the number of total layers, $d_{model}$ denotes embedding size, $n_{head}$ denotes the number of attention heads.}\label{tb:scaling-model}
\begin{tabular}{@{}rccccc@{}}
\toprule
\textbf{\makecell{\#Paras \\ (non-emb)}}     & \textbf{$n_{layer}$}    & \textbf{\makecell{$d_{model}$}}     & \textbf{$n_{head}$}     &\textbf{Epochs}     & \textbf{\makecell{Batch Size \\ (global)}}\ \  \\ \midrule
98,304           & 2 & 64  & 2  & 30 & 256\ \ \    \\
786,432            & 4 & 128  & 4 & 30 & 256\ \ \    \\
1,572,864            & 8 & 128  & 4 & 30 & 256\ \ \    \\
9,437,184            & 12 & 256  & 8   & 27 & 256\ \ \    \\
75,497,472            & 24 & 512  & 8 & 12 & 256\ \ \    \\
829,440,000 & 48  & 1200  & 24  & 12 & 256\ \ \     \\ \bottomrule
\end{tabular}
\end{table}

\subsection{Performance Measurement} 
In our work, we consider two main kinds of performance measures for large SR models, namely the ID-based modeling measure and the evaluation metrics for recommendation tasks. 

For ID-based modeling measure, we adopt 
{the cross-entropy loss as the performance measurement for scaling. Specifically, each item is treated as a separate category, and the model calculates the probability that the next item belongs to each category. The loss will be averaged over the items in a sequence}.

For recommendation evaluation, we group the interactions for each user and sort them chronologically following previous studies~\cite{kang2018self,sun2019bert4rec,fan2021lighter}. We adopt the widely-used leave-one-out strategy, in which the last item is used as the test item, the item before the last item is used as the validation item, and the remaining items are used for training. For each dataset, we filter all test items that don't appear in training and validation datasets. In terms of evaluation metrics, we utilize Hit Ratio@N (HR@N) and Normalized Discounted Cumulative Gain@N~(NDCG@N) for accuracy evaluation and Item Coverage~(Coverage) for diversity evaluation, where N $\in$ \{5, 10, 50\}.

\begin{figure*}[h]
	\centering
	\begin{minipage}{1.0\textwidth}	
            \hspace{0.3in}
	    \subfigure[Model Scaling Curve]{
		  	\label{fig:model_scaling}
			\includegraphics[width=0.4\textwidth]{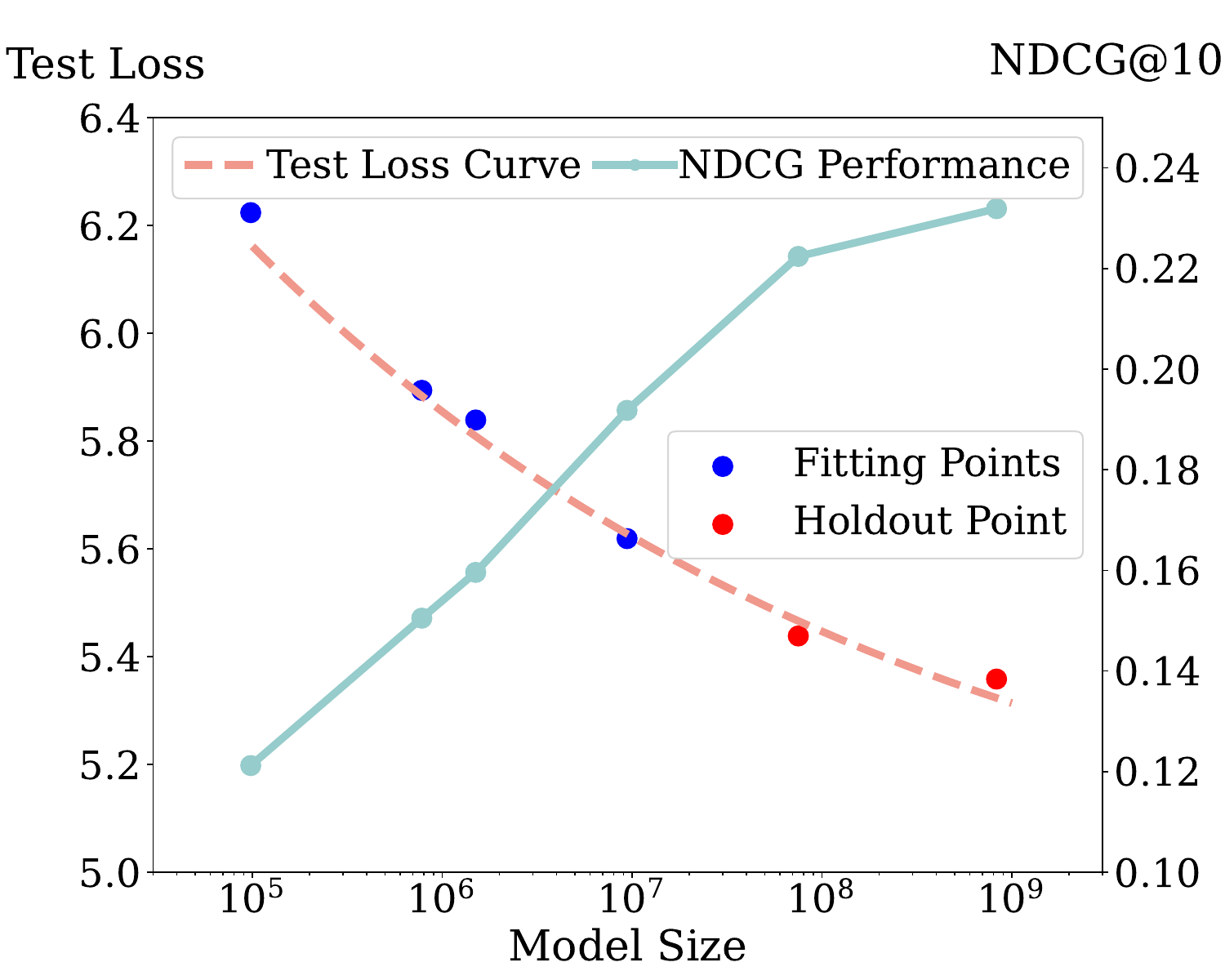}	
		}\noindent
            \hspace{0.5in}
		\subfigure[Data Scaling Curve]{
		    \label{fig:data_scaling}
		    \includegraphics[width=0.4\textwidth]{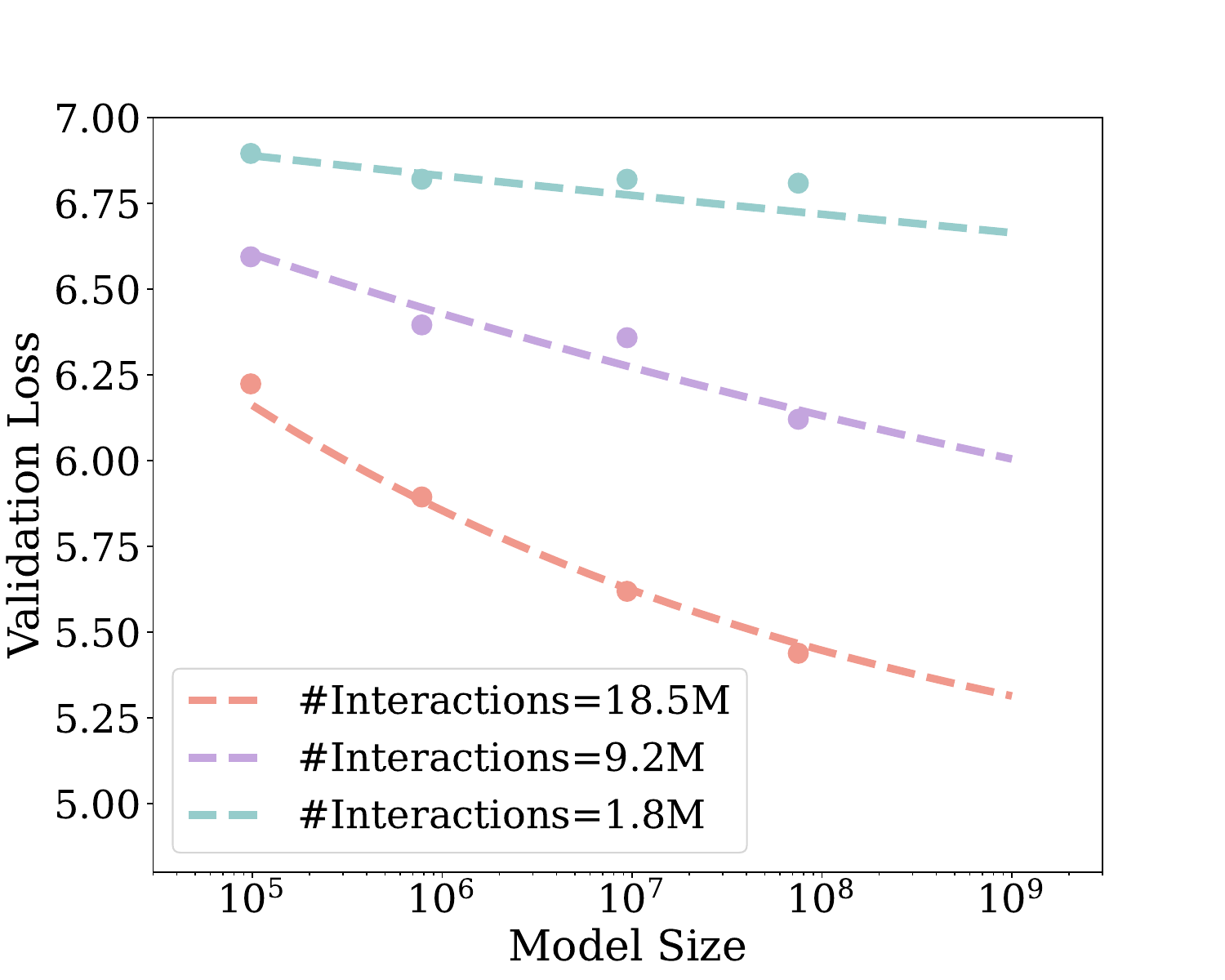}
		}
	\end{minipage}
	
	\caption{Scaling Curve on Model Size and Data Size on MovieLens dataset.}
	\label{fig:ml_scaling}
\end{figure*}

\subsection{Efficiency Analysis}

Efficiency is a crucial evaluation factor for recommendation models, particularly for large-scale models. To assess their practical deployment, we further analyze the time and space complexity of the model in this section. In each layer, the calculation of self-attention matrices is typically the most computationally expensive operation, with a complexity of $O(n^2d)$, where $n$ is the sequence length and $d$ is the embedding size. Therefore, for an $L$-layer model, its time complexity is $O(n^2dL)$. For space complexity, model parameters mainly come from item embedding, self-attention blocks, feed-forward network and layer normalization. Therefore, its space complexity is $O(|\mathcal{I}|d + nd + d^2)$, where $\mathcal{I}$ is the total item set. 

Based on this analysis, it may be challenging to use large-scale models due to their large memory footprints and high latency. However, both the large memory footprints and high latency are determined by the total number of bits used in model parameters. To address this issue, model quantization may be an effective way to reduce model bits. Model quantization is a compression technique that transforms floating-point storage (operations) into integer storage (operations). This technique involves converting the original model weight from FLOAT32 to INT8 or even INT4, thereby leading to faster inference acceleration and substantial savings in model storage. Although it has been shown to achieve inference acceleration and model storage reduction without significant performance degradation, its effectiveness for recommendation models remains underexplored. This aspect of research is left for future work. 
For our experimental setups, all models are trained/tested on a Linux machine with an AMD CPU, 1T memory and eight NVIDIA A100 40GB GPUs. 

\section{SCALING LAWS in SEQUENTIAL RECOMMENDATION}\label{sec:scaling}

In this section, we investigate the scaling laws for sequential recommendation performance (measured by cross-entropy loss). As discussed in prior work~\cite{kaplan2020scaling,henighan2020scaling}, there are generally three key factors that are considered to affect scaling properties of models~\cite{kaplan2020scaling,henighan2020scaling}: compute $C$, model size $N$, and data size $D$. 
In our experiments, we mainly focus on examining the influence of model size $N$ and data size $D$. To ensure accurate results, we maintain adequate compute $C$ throughout our experiments. Additionally, we also explore other relevant factors that potentially affect the scaling effect, such as the model shape. Next, we present the main experiments conducted and the corresponding results obtained.

\subsection{Scaling Properties on Model Capacity}\label{subsec:model-scaling}

We first examine the influence of scaling model capacity, \ie the \textit{non-embedding} trainable parameters of Transformers~\cite{kang2018self}. We keep $d_{\mathrm{model}} = d_{\mathrm{attn}} = d_{\mathrm{ff}} \, / \, 4 $ as the standard setup in this section ($d_{\mathrm{ff}}$ here denotes $d_{\mathrm{feedforward}}$). The specific configurations for model capacity scaling are presented in Table~\ref{tb:scaling-model}. 

In order to quantitatively measure the impact of scaling model capacity, we introduce a power-law function to be fitted. Specifically, we treat the non-embedding parameters $N$ as independent variables, while $E_N$, $N_0$, $\alpha_{N}$ are constants to fit for each dataset. Here $E_N$ is the irreducible loss which estimates the entropy of the underlying data distribution, and $\alpha_{N}$ is the dataset-dependent scaling exponent. The scaling law of test loss (single epoch) on model capacity can be summarized as $L(N)$: 

\begin{equation}
\label{eq:scaling}
L(N) = E_N + (N_0 / N)^{\alpha_{N}} 
\end{equation} 

For the MovieLens-20M dataset, by fitting using four models ranging from 98.3K to 9.4M (the blue dots in Figure~\ref{fig:model_scaling}), we have the $E_N=4.9$, $N_0=6.8\times 10^5$, $\alpha_{N}=0.121$ and the power-law curve (the red dot-line). As shown in Figure~\ref{fig:model_scaling}, we observe that the test loss at a single epoch follows a power-law relationship with the number of non-embedding parameters, while keeping the data size fixed.
This implies that increasing model capacity can benefit the performance of recommendation models. 

Furthermore, compared with the $\alpha_{N}=0.07$ in language models, we can find that $\alpha_{N}$ is relatively larger in sequential recommendation. This suggests that the decrease in loss with model capacity is slightly faster in sequential recommendation compared to NLP. Our results suggest that in sequential recommendation, the potential benefits of scaling up model size may be greater than those observed in NLP, if not limited by data. 
Additionally, the curve changes for different datasets. We find that $\alpha_{N}$ decreases as the data sparsity increases. As the number of interactions is reduced and the data becomes sparser in Figure~\ref{fig:data_scaling}, the power-law curve gradually becomes flatter, indicating that $\alpha_{N}$ becomes smaller. This suggests that the slower decrease in the test loss on some datasets may be due to data sparsity.

We further utilize the fitted power-law curve to predict the performance of two larger models. As shown in Figure~\ref{fig:model_scaling}, the red dots represent the actual performance of two large models~(75.5M and 0.8B), and they align closely with the prediction curve. This finding implies that predictable scaling can be explored for recommendation models, where the performance of small models'~(\textless 100X) can accurately predict large models' performance. Additionally, we observe that the model performance continues to increase when scaling up the model capacity exceeding 0.8B parameters. 

In addition to evaluating the model performance using test loss, we also measure its performance on the recommendation task using Normalized Discounted Cumulative Gain (NDCG) and observe similar scaling laws. However, we notice low-return regimes in the NDCG curve at 0.8B parameter level. We speculate that the model performance might be still limited by data size, and further explore the relationship between model performance and data size in the next subsection.

\subsection{Scaling Properties on Data Size}\label{subsec:data-scaling}

In addition to model size, data size is also an important factor to consider for scaling law. In particular, recommender systems have usually different data distributions compared to other domains, such as high data sparsity.

We first analyze the impact of data scaling on model performance by considering different training datasets of varied sizes. Specifically, we create datasets consisting of 1.8M, 9.2M, and 18.5M interactions and keep global batch size constant during training. For each data size level, we train models of four sizes (98.3K, 0.78M, 9.44M, and 75.5M trainable parameters) and present their test loss in Figure~\ref{fig:data_scaling}. The scaling law curves are fitted for each data size level, and imply several key findings: 

\textbf{Scaling law holds when data size varies, including in highly constrained cases.} In recommender systems, data tends to be more sparse compared to other domains such as NLP, leading to a \textit{data-constrained setting}. For instance, the largest model in our experiments has 0.8B parameters, while the available dataset only has 18.5M interactions. In contrast, a language model of the same size usually requires a much larger dataset (more than 10B tokens)~\cite{kaplan2020scaling}. 

In such a data-constrained scenario, the scaling law still holds in sequential recommendation, even when we further decrease the data size to 1.8M (only 10\% of the whole dataset). In different data size levels, we observe that the test loss can be fitted with a power-law function, consistent with the findings in Section~\ref{subsec:model-scaling}, as shown in Figure~\ref{fig:data_scaling}. 

\textbf{Larger models are more data-efficient.} As shown in the curves in Figure~\ref{fig:data_scaling}, we find that larger models can achieve lower losses even with small datasets, while smaller models require a larger amount of data to achieve similar loss values. For instance, the model with 75.5M parameters can achieve a test loss of 6.1217 on 9.2M data, while the model with 98.3K parameters requires 18.5M data to achieve the same level of loss. Furthermore, by investigating the benefits obtained by the models of different sizes as the amount of data increases, we find that the larger the model, the larger the improvement from the same increase in data size. This indicates the higher data utilization efficiency of larger models.



\begin{figure}[t]
    \centering
    \includegraphics[width=0.44\textwidth]{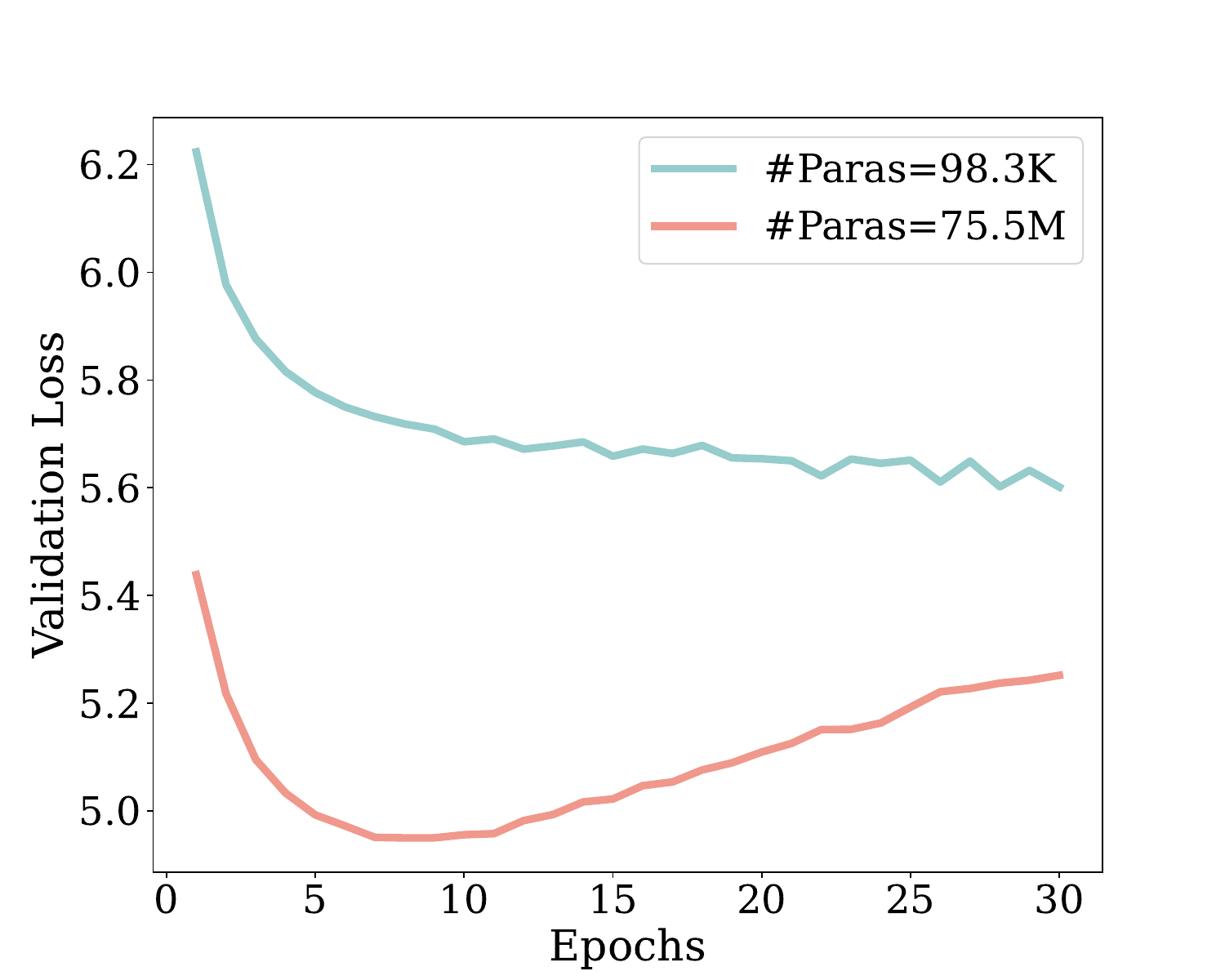}
    \caption{Test loss with data repetition.} 
    \label{fig:loss}
\end{figure}

\subsection{Scaling with Data Repetition}\label{subsec:repeat}

Scaling with a shortage of training tokens is often called \emph{data-constrained scaling}~\cite{muennighoff2023scaling}, where it is difficult to reach convergence by iterating one pass over the entire dataset. In the recommendation scenario, interaction data is usually insufficient compared to its large-scale users and items, even in some large-scale real-world industry scenarios. In language modeling, repeating data has been utilized to improve the performance when scaling under a data-constrained regime~\cite{muennighoff2023scaling}. 

In this subsection, we investigate whether data repetition can bring benefits in scaling recommendation models, alleviating the data-constrained issue. Note that we have selected two large recommendation datasets, the MovieLens-20M and the whole Amazon-2018 (without domain split), but the number of interactions is still only 18.5M and 50.6M, smaller than the model parameters. It is difficult to reach complete model convergence in one single data epoch. We repeat the training samples in the datasets, \ie iterating the optimization in multiple epochs. 

Specifically, we train a large model~(75.5M parameters) and a small model~(98.3K parameters) for up to 30 epochs respectively, and record the cross-entropy loss on the validation dataset. As shown in Figure~\ref{fig:loss}, we can find that neither the 75.5M model nor the 98.3K model reached complete convergence at one epoch, while they all benefit from additional epochs~(2-5 epochs). It indicates that data repetition is important for the insufficient data issue in recommender systems. Moreover, we observe rapidly diminishing returns for more repetitions~(6-12 epochs), which implies that the information in the data is gradually learned by models. Finally, the returns eventually diminish to zero~(13-30 epochs) in both models and the large model exhibits an increasing risk of overfitting, as the model performance is limited by the size of unique data and repeating is worthless.

\begin{figure}[t]
    \centering
    \includegraphics[width=0.44\textwidth]{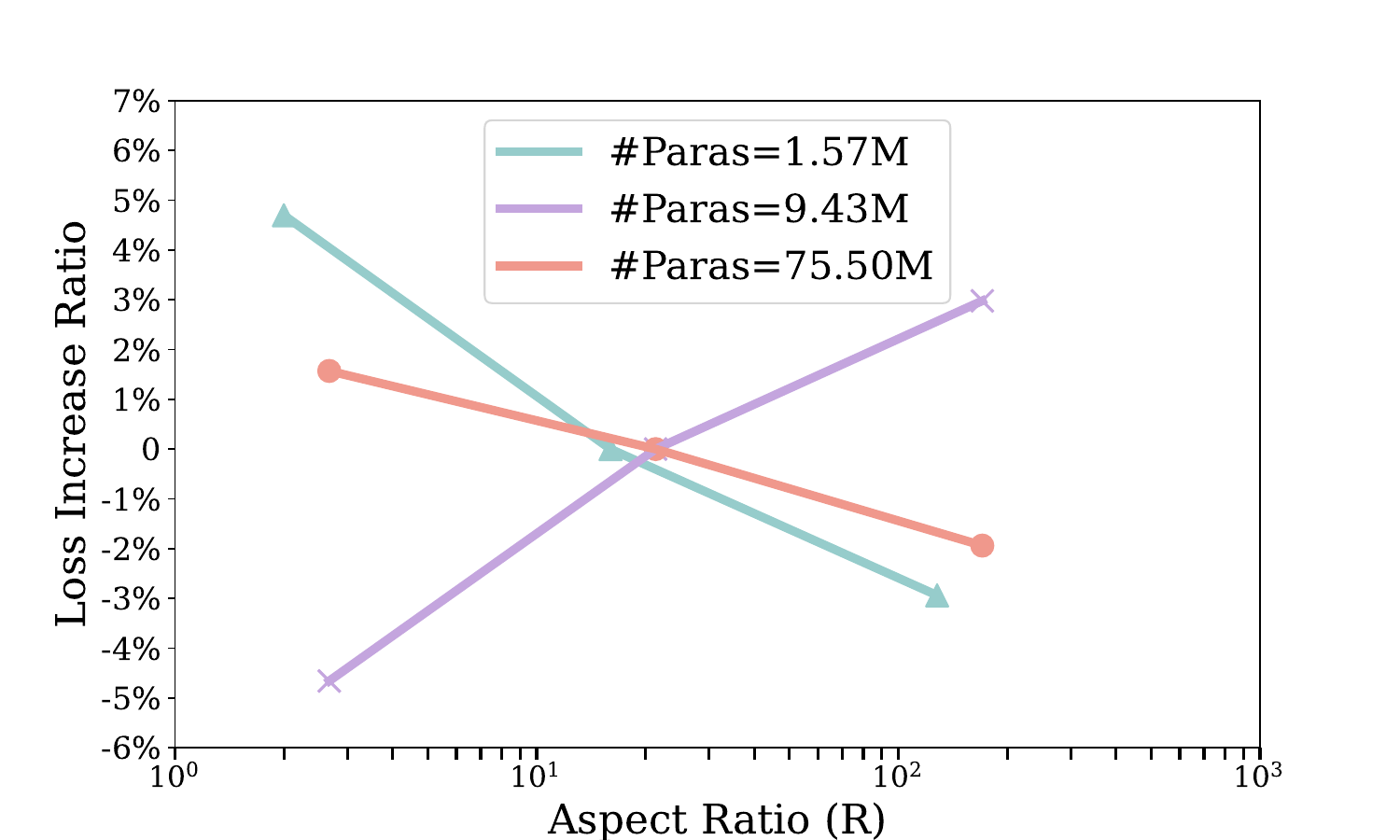}
    \caption{Loss increase with different aspect ratios and model scales. Each point represents a shape transformation, and the X-axis ``Aspect Ratio'' is used to measure the specific transformation. Loss increase in the Y-axis refers to the percentage loss increase caused by the shape transformation compared to the standard shape in Table~\ref{tb:scaling-model}.} 
    \label{fig:aspect}
\end{figure}

\subsection{Effect of Model Shape on Scaling}\label{subsec:shape}

Determining which part of the model to scale when increasing its size is an important consideration. The model shape is typically characterized by two major factors, model depth~($n_{layer}$) and model width~($d_{model}$). For a fixed total parameter $N$, a greater model depth means a deep and thin model; a greater model width means a wide and shallow model. Kaplan \emph{et al.}~\cite{kaplan2020scaling} found model performance depends mildly on model shape in language modeling, but there is little understanding in recommender systems.

To gain further insight into the impact of model shape, we examine the relationship between model performance and the transformer model shape. We measure the model shape using the \textit{aspect ratio} $R$~\cite{kaplan2020scaling} which is calculated as:
\begin{equation}
    R = d_{model} \, / \, n_{layer}
\end{equation}
It can well reflect the ratio between the depth and width of the model, indicating a deeper or wider model shape.

Following~\cite{kaplan2020scaling}, we simultaneously vary different hyper-parameters meanwhile keeping total \textit{non-embedding} parameters $N$ fixed. We train models of three sizes: 1.57M, 9.43M, and 75.5M. For each model size, we keep $N$ fixed and perform three shape transformations, recording the loss of each transformation. As an example, for the smallest model size of 1.57M, the aspect ratio values for the three transformations are 2~($n_{layer}=32,d_{model}=64$), 16~($n_{layer}=8,d_{model}=128$), and 128~($n_{layer}=2,d_{model}=256$), covering a wide range of selection. Figure~\ref{fig:aspect} shows how the loss value changes for each transformation. The standard shape in Table~\ref{tb:scaling-model} is considered as the baseline. We observe that even when we vary the model shape over a wide range, the increase in loss for each shape transformation is minimal, indicating that model performance exhibits weak dependence on the model shape. Furthermore, as the model size gradually increases, this influence tends to decrease further, as shown by the three curves in Figure~\ref{fig:aspect}.

\begin{table*}[t]
\centering
\renewcommand{\arraystretch}{1.5}
\caption{Overall performance on recommendation. LSRM denotes our Large Sequential Recommendation Model, \#Paras~(emb) denotes total embedding parameters, \#Paras~(non-emb) denotes total \textit{non-embedding} parameters.}\label{tb:overall}
\begin{tabular}{@{}lcccrrcccccc@{}}
\toprule
    & \multicolumn{5}{c}{\textbf{Model Strcture}} & \multicolumn{6}{c}{\textbf{Performance}}\\
    \cmidrule(r){2-6}\cmidrule(l){7-12}
     & \textbf{$n_{layer}$}    & \textbf{$d_{model}$}    & \textbf{$d_{ff}$}   & \textbf{\makecell{\#Paras \\ (emb)}}     & \textbf{\makecell{\#Paras \\ (non-emb)}}  & \textbf{HR@5}    & \textbf{NDCG@5}  & \textbf{HR@10}    & \textbf{NDCG@10}  & \textbf{HR@50}  & \textbf{NDCG@50}\ \  \\ 
\Xhline{0.5pt}

SASRec    & 2   & 64   & 256   & 1.69M & 98.3K & 0.1463 & 0.093 & 0.2337 & 0.1212 & 0.5166 & 0.1834\ \ \    \\
GRU4Rec       & 2   & 64   & -     & 1.69M & 180.3K & 0.1559 & 0.1007  & 0.2466 & 0.1281 & 0.5267 & 0.1965\ \ \    \\
FMLP          & 2   & 64   & -  & 1.69M & 167.6K & 0.1498 & 0.0978 & 0.2427 & 0.1255 & 0.5214 & 0.1918\ \ \    \\
Caser         & 2   & 64   & -   & 10.55M & 703.7K & 0.1407 & 0.0877 & 0.2219 & 0.1140 & 0.5036 & 0.1793\ \ \    \\
${\mathrm{LSRM}}_{wide}$   & 2   & 1200   & 4800         & 31.77M & 34.56M & 0.0595 & 0.0384 & 0.0976 & 0.0506 & 0.2639 & 0.0865\ \ \    \\
${\mathrm{LSRM}}_{deep}$ & 48   & 1200   & 4800  & 31.77M  & \textbf{0.83B}   & \textbf{0.2794} & \textbf{0.1827} & \textbf{0.3786} & \textbf{0.2319} & \textbf{0.6264} &\textbf{0.2794}\ \ \     \\ \bottomrule
\end{tabular}
\vspace{-0.15in}
\end{table*}

\section{BENEFITS OF LARGE RECOMMENDATION MODELS}\label{chap:5}

In Section~\ref{sec:scaling}, we examine the scaling laws of sequential recommendation models on general performance. In this section, we delve deeper into the potential advantages of large-scale recommendation models across various downstream tasks, as general test loss is not always the comprehensive measure of recommendation performance in real-world applications. First, we evaluate the performance of different scale models in conventional recommendation scenarios. In addition, we speculate and empirically verify that larger models tend to be more capable of solving more difficult recommendation tasks, including \textit{long-tail item} recommendation, \textit{cold-start user} recommendation, robustness challenge against \textit{noisy inputs}, \textit{multi-domain} transfer, and \textit{user trajectory prediction}.


\subsection{Overall Performance}\label{subsec:overall}

Besides test loss, we further evaluate the scaling models on the sequential recommendation task. Specifically, we follow the conventional benchmark settings, and split the data into training, validation and test sets using the \textit{leave-one-out} strategy~\cite{zhou2022filter}. Note that our observations in Section~\ref{subsec:repeat} have revealed that the utilization of repeated data can yield additional benefits when data is highly constrained. Consequently, all models presented in this section have been trained until convergence with data repetition.


For comparison, we include various state-of-the-art sequential recommendation models as baselines, such as SASRec, GRU4Rec~\cite{hidasi2015session}, FMLP~\cite{zhou2022filter}, and Caser~\cite{tang2018personalized}. As for the scaled Large Sequential Recommendation Model (namely LSRM), we train two variants, $\mathrm{LSRM}_{wide}$ and $\mathrm{LSRM}_{deep}$. The number of total parameters of $\mathrm{LSRM}_{deep}$ is $\times$8k times larger than the default SASRec model and $\times$1k times larger than the default Caser model. As shown in Table~\ref{tb:overall}, we find that $\mathrm{LSRM}_{deep}$ significantly outperforms the baselines, being almost two times better on the HR@5, NDCG@5 and NDCG@10 metrics. This shows that scaling up model size may bring greater improvements than changing the model structure. Moreover, a comparison between $\mathrm{LSRM}_{deep}$ and $\mathrm{LSRM}_{wide}$ reveals that solely augmenting the model width led to a substantial decline in performance, aligning with the embedding collapse phenomenon in prior work~\cite{guo2023embedding}. This observation further highlights the necessity of exploring the scaling properties of \textit{non-embedding} parameters within transformers.


In the following, we examine the performance of LSRM models on five more complex recommendation tasks. These tasks are either the long-standing challenges of recommender systems, or problems with real-world needs. Surprisingly, large recommendation models show great potential in all of these tasks, as we will illustrate in the following subsections.

\begin{figure*}[t]
	\centering
	\begin{minipage}{1.0\textwidth}	
            \hspace{0.2in}
		\subfigure[MovieLens-20M]{
			\label{fig:ml_pop}
			\includegraphics[width=0.4\textwidth]{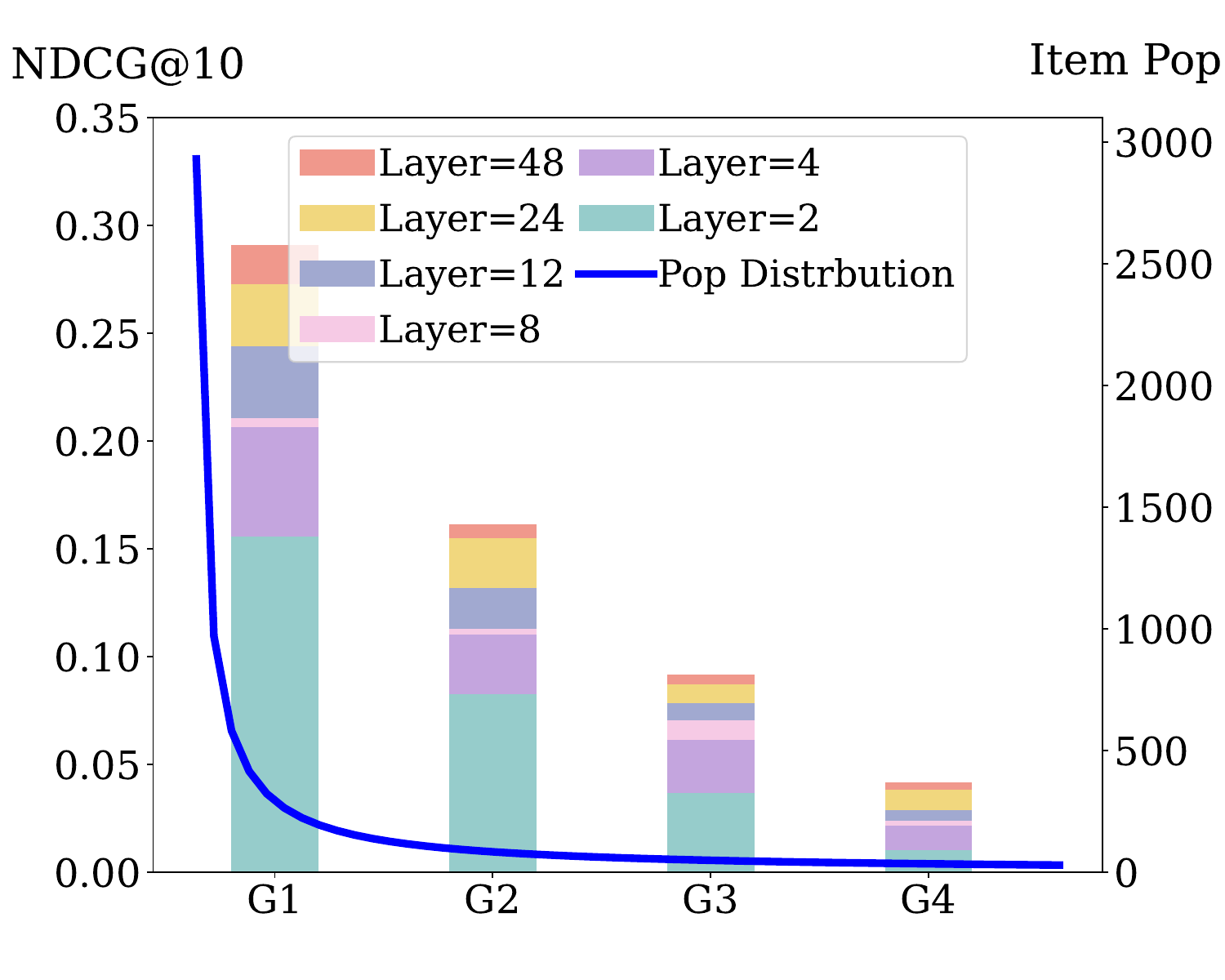}	
		}\noindent
            \hspace{0.4in}
		\subfigure[Amazon]{
			\label{fig:amazon_pop}
			\includegraphics[width=0.4\textwidth]{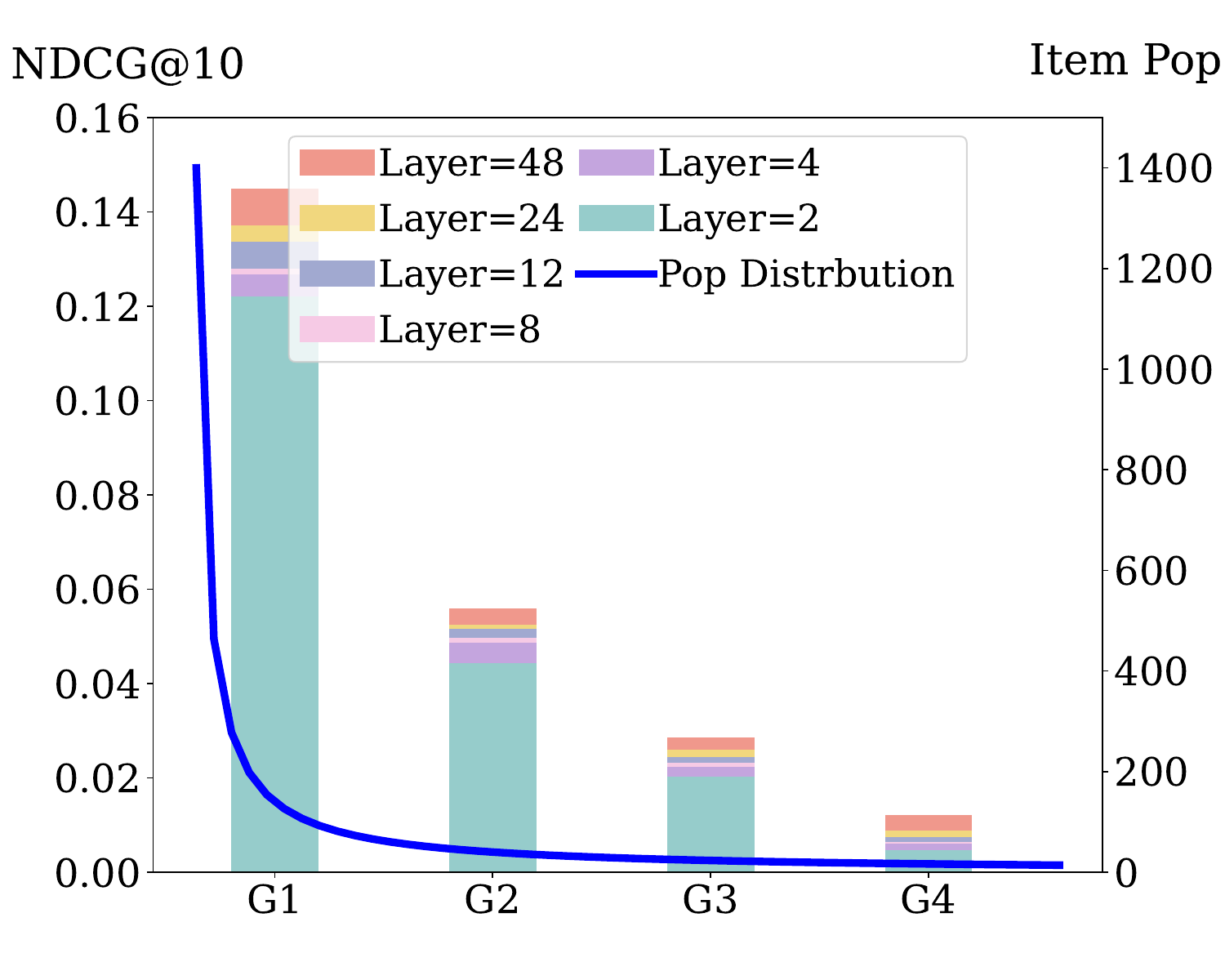}
		}
	\end{minipage}
	
	\caption{Popularity distribution of items and performance of models at different scales on different popularity groups. G1 denotes the group of target items with the highest average popularity.}
	\label{fig:item_pop}
\end{figure*}

\subsection{Long-tail Item Recommendation}

In recommender systems, the distribution of item popularity (\eg measured by interaction frequency) tends to be skewed and imbalanced. 
These infrequent items (often called \emph{long-tail items}) receive little attention. It is usually challenging for recommendation models to effectively recommend long-tail items due to a lack of data and insufficient learning. 
Recent evidence has shown that models can generalize better on few-shot examples by scaling up the model size~\cite{brown2020language,openai2023gpt4,hoffmann2022training}. In this subsection, we explore whether LSRMs can alleviate the long-tail problem.


We sort the items in descending order based on their popularity (\ie the interaction number) and split them into four groups of different popularity. The item popularity distributions on both MovieLens and Amazon datasets are clearly long-tailed, as shown in Figure~\ref{fig:item_pop}~(blue line). The four groups of items are equal-sized to ensure a fair comparison. After that, we evaluate the model performance with varying model scales on each group(bars). 

Firstly, we find that large models consistently outperform small models in each popularity group, showing that the improvements brought by scaling are stable. Secondly, we can also observe that as the item popularity decreases, the performance gap between the large and small models significantly increases. For instance, in the group G1 with the highest average popularity, the 48-layer large model (${\mathrm{LSRM}}_{l_{48}}$) achieves twice the performance of the 2-layer small model (${\mathrm{LSRM}}_{l_2}$). While in the group G4 with the lowest average popularity, ${\mathrm{LSRM}}_{l_{48}}$ can achieve 3 times the performance of the ${\mathrm{LSRM}}_{l_2}$. This observation suggests that large-scale LSRMs achieve additional benefits when dealing with long-tail item recommendations.
The reason can be twofold. First, large models have shown strong generalization ability on few-shot examples and tasks~\cite{brown2020language,hoffmann2022training,openai2023gpt4}. As a result, LSRMs have higher chances to generalize the strong recommendation capabilities to long-tailed items, even when these items are only associated with a few training cases. Second, large models are believed to have better memorization ability~\cite{carlini2022quantifying,tirumala2022memorization}. In this way, the limited training cases may be well memorized by large recommendation models, and will not be overwhelmed by those interactions with popular items.

\begin{figure*}[t]
	\centering
	\begin{minipage}{1.0\textwidth}	
            \hspace{0.2in}
	    \subfigure[MovieLens]{
		  	\label{fig:ml_length}
			\includegraphics[width=0.4\textwidth]{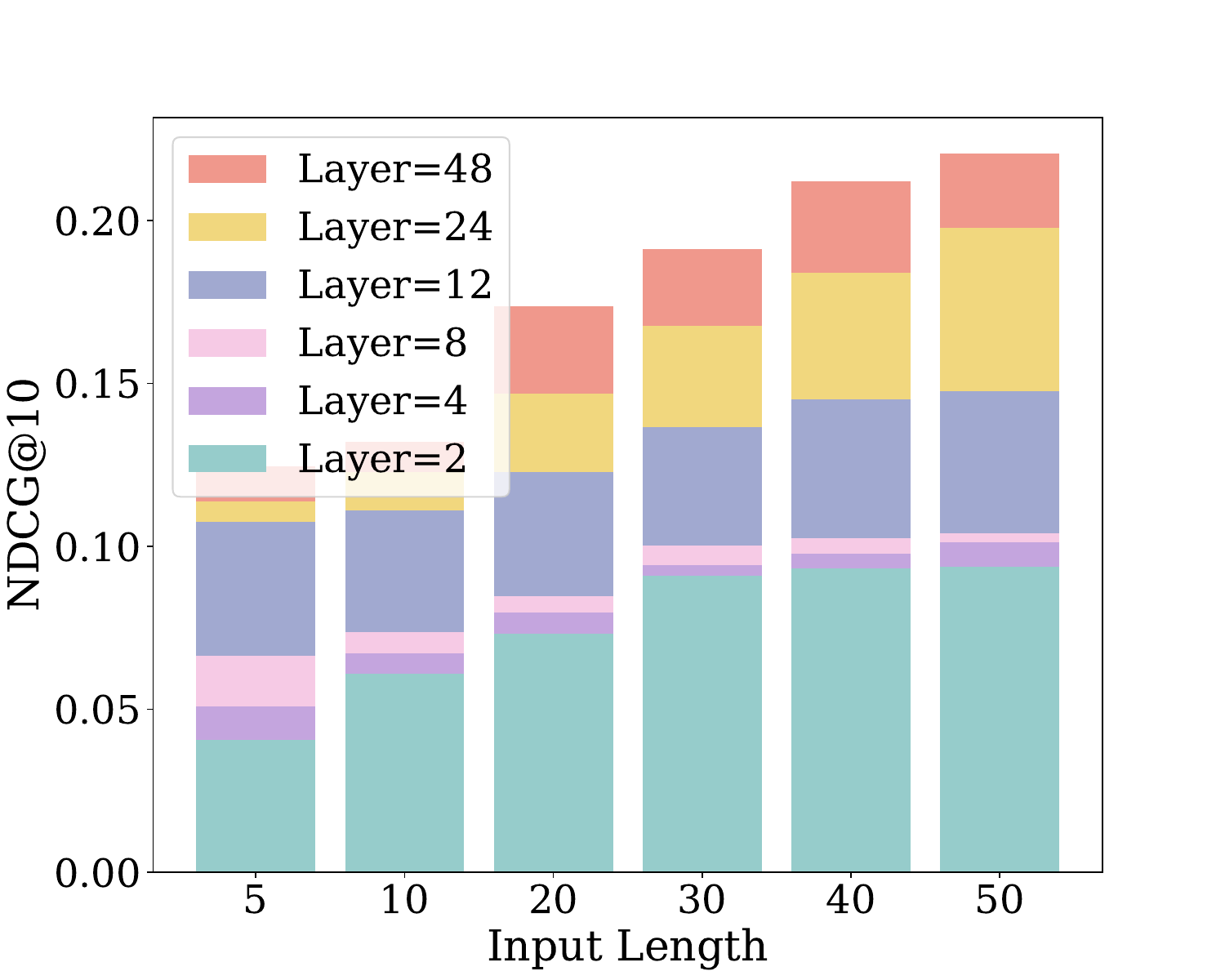}	
		}\noindent
            \hspace{0.4in}
		\subfigure[Amazon]{
		    \label{fig:amazon_length}
		    \includegraphics[width=0.4\textwidth]{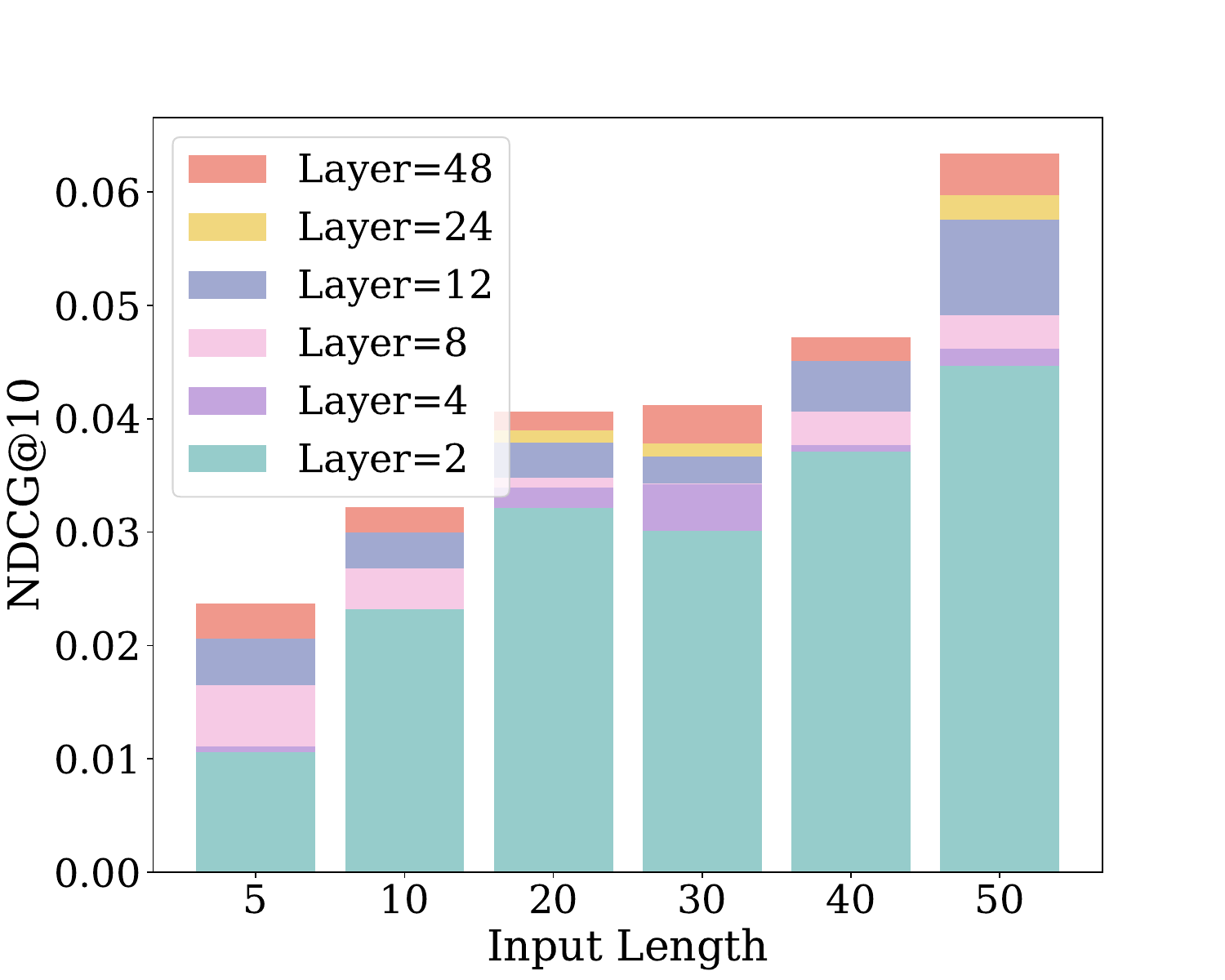}
		}
	\end{minipage}
	\caption{Model performance on different input length groups at different scales.}
	\vspace{-0.15in}
	\label{fig:historical}
\end{figure*}

\begin{figure*}[t]
\centering
\begin{minipage}{1.0\textwidth}	
    \hspace{0.2in}
    \subfigure[Numerical comparison]{
		\label{fig:cross_num}
		\includegraphics[width=0.4\textwidth]{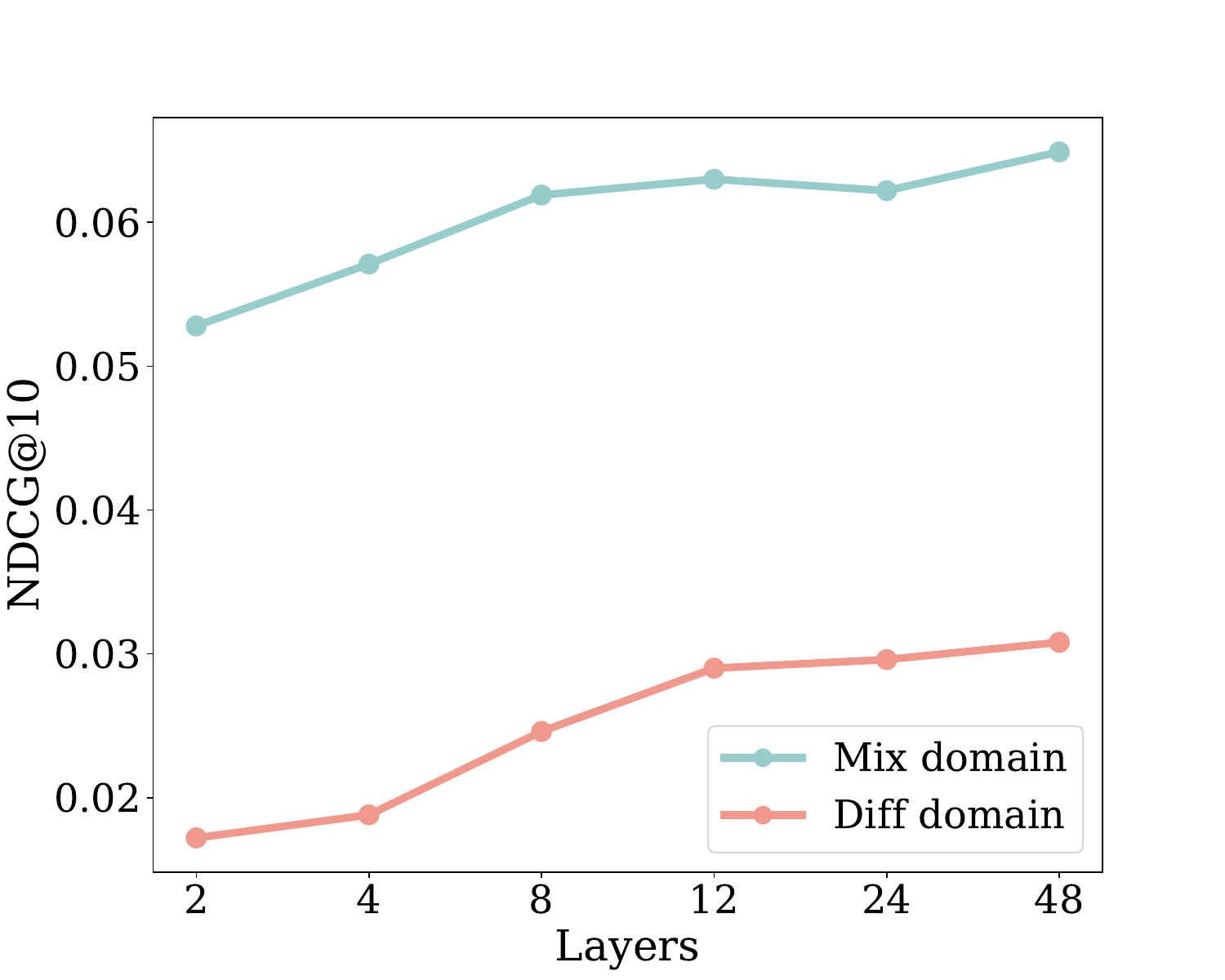}	
	}\noindent
    \hspace{0.4in}
	\subfigure[Percentage comparison]{
		\label{fig:cross_percentage}
	    \includegraphics[width=0.4\textwidth]{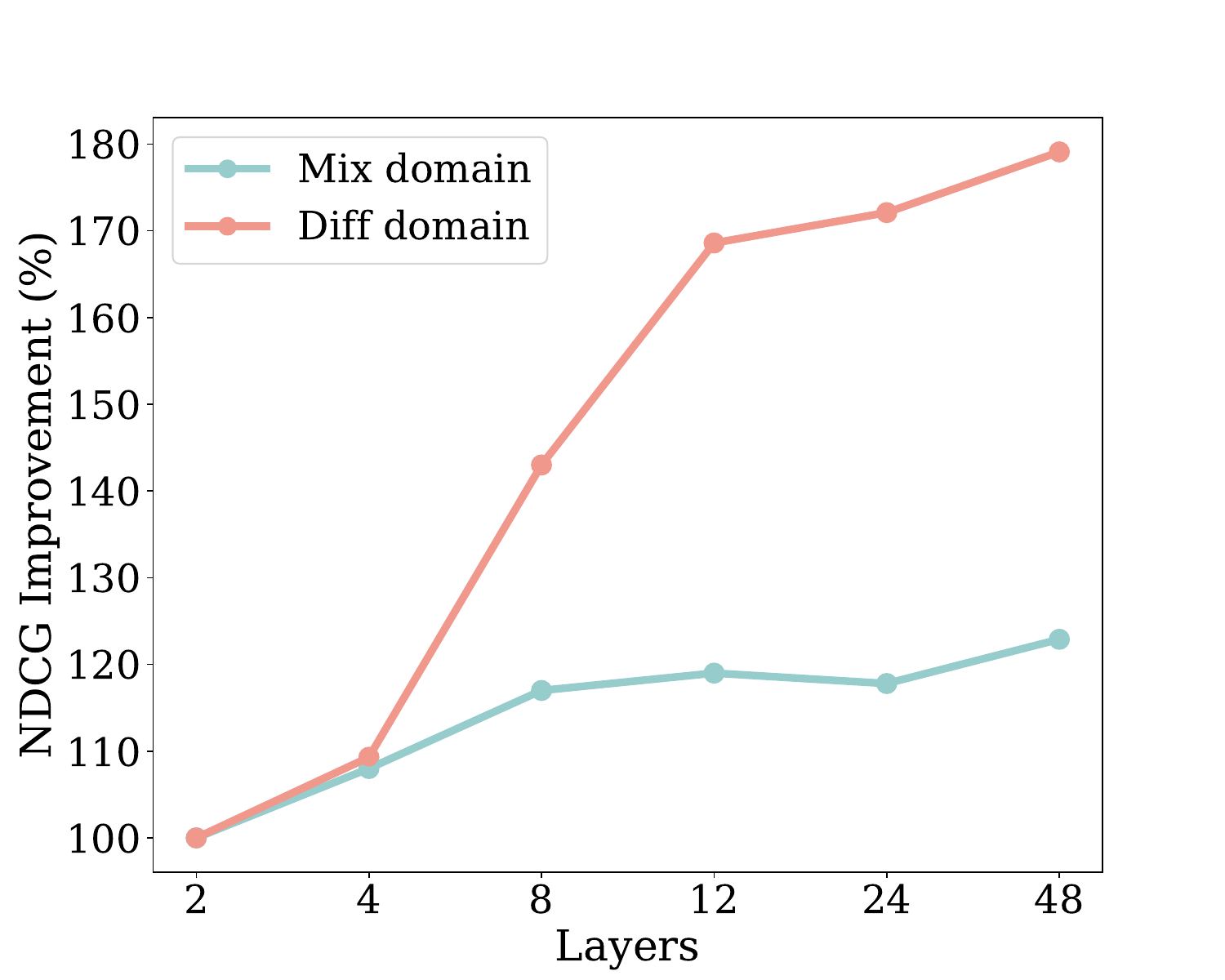}
	}
\end{minipage}
	
\caption{Performance comparison of different scale models in the mix-domain recommendation and diff-domain recommendation.}
\label{fig:cross}
\end{figure*}

\subsection{Cold-start User Recommendation}

In addition to long-tail item recommendation, another long-standing challenge in recommender systems is cold-start user recommendation, which also arises due to data sparsity. Since the key part of the recommendation is personalization, it is significantly difficult to capture user preference from limited user interactions accurately. In this subsection, we explore whether LSRMs can better model user preferences from insufficient interactions. 

To investigate this, we conduct experiments where we extract and examine historical interactions of different lengths. Specifically, we randomly select 20\% of the users from the dataset as new users who did not participate in the training and are used for testing. We simulate the cold-start scenarios of different levels by utilizing only a portion of their historical sequences. For instance, to simulate a new user scenario with only five interactions, we keep the last five interactions of users, and evaluate the model performance on the leave-one-out test set. We examine the performance of LSRMs across different input lengths. 

Figure~\ref{fig:historical} shows the NDCG performance of models of various sizes with different input lengths. We observe a clear trend: as the input length increases from 5 to 50, larger models consistently maintain a performance advantage over smaller models. Interestingly, we find that the performance improvements of larger models are more significant when dealing with extremely short input sequences.
One possible reason is that large-scale models can provide more diverse recommendation results for cold-start users, which are more likely to match the users' interests. We examine the recommendation diversity~(using Coverage@10) of the cold-start user group~(input length=5) on the MovieLens-20M dataset. For the 2-layer small model, the Coverage@10 value is 0.1362, while for the 48-layer large model, its Coverage@10 value is 0.2046. This finding verifies our conjecture to a certain extent.

\subsection{Transfer on Multi-domain Recommendation}

In the context of sequential recommendation, multi-domain transfer is also a challenging issue in real-world applications. Here, we consider a commonly studied multi-domain sequential recommendation~(MDSR) setting~\cite{cao2022contrastive,tang2023one}. For each user $u$, we collect his/her interactions from all domains and form a mixture sequence chronologically, where items in the sequence may come from different domains. Then given the mixed interaction sequences, MDSR aims to predict the next item in a target domain $D^T$.
Furthermore, we also conduct experiments to explore how models of different scales perform on multi-domain transfer.

Specifically, we experimented with cross-domain recommendations on the Amazon~\cite{ni2019justifying} dataset. As introduced in Section~\ref{subsec:train-data}, we have already mixed the sequences from all domains on this dataset. Following~\cite{tang2023one}, we further refine the multi-domain setup based on the relationship of the target item domain and sequence item domain as below:

\begin{itemize}
    \item \textbf{Mix-domain}: Some items in the sequence belong to the same domain as the target item, while the rest are from other domains.
    \item \textbf{Diff-domain}: The domains of all items in the sequence are distinct from the domain of the target item.
\end{itemize}
To investigate the impacts of model scales on this multi-domain recommendation transfer task, we conduct experiments with varying model sizes. It should be noted that we only classify the test set based on the above rules, and the models are all trained on mixed sequences in Amazon.

The results of our experiments are depicted in Figure~\ref{fig:cross}. As shown in Figure~\ref{fig:cross_num}, we find that the curve of `Mix-domain' is always above the `Diff-domain' curve, which indicates that `Diff-domain' is a more difficult task than `Mix-domain'. This demonstrates the difficulty of transferring user preferences to a completely different domain. In addition, we use ${\mathrm{LSRM}}_{l_2}$ on two tasks as baselines to evaluate the improvement of the models at each scale. Upon analyzing the percentage improvement results in Figure~\ref{fig:cross_percentage}, we observe a notable trend: as the model size increases, the performance improvement of the model in the `Diff-domain' setting far exceeds that in the `Mix-domain' setting. It highlights the advantages of large-scale models in multi-domain knowledge transfer. 

\subsection{Robustness Challenge}

In recommender systems, interaction data is usually more noisy since it is generated by free user interaction behaviors. For example, a click record might be triggered by an occasional behavior of a user, rather than a reliable signal of user interest. It is important to enhance the model robustness against these noises in sequential recommendation. Many studies~\cite{tan2023towards,he2018adversarial} try to improve the model robustness by applying adversarial perturbations to item interactions or embeddings during the training process. Here, we aim to investigate whether the intrinsic capacity to resist noise can be enhanced if we scale up the model size.  


According to the literature~\cite{o2004collaborative,o2005trust,tan2023towards}, model robustness is often studied by examining how well the recommendation results hold up against various perturbations applied to the input sequences or item embeddings. To simulate real-world recommendation scenarios, we investigate the impact of perturbations directly applied to the sequence on the model. Specifically, perturbations in our experiment include \emph{removing}, and \emph{replacing} items in the input sequence. Each item has an equal probability of being perturbed.

\begin{figure}[t]
\centering
\includegraphics[width=0.44\textwidth]{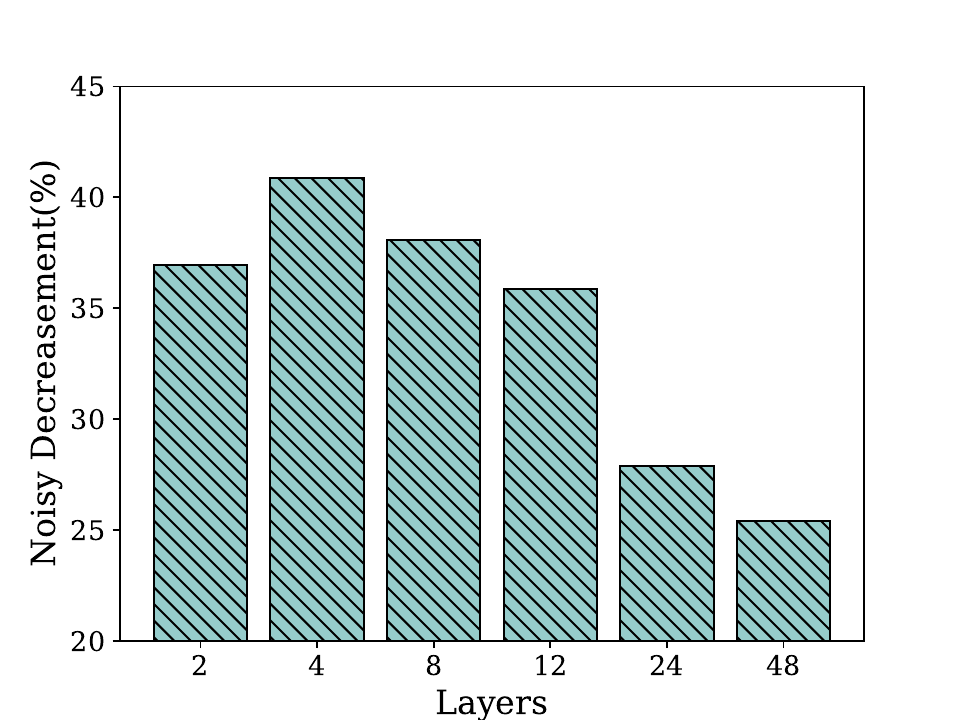}
\caption{Percentage performance degradation against noisy input sequences at each scale.}
\label{fig:robustness}
\end{figure}

By rearranging the test sequences with different perturbations, we simulate changes in user behaviors and test the model's stability in varied contexts. The performance degradation of LSRMs with different sizes is shown in Figure~\ref{fig:robustness}. Although all models suffer performance degradation in the presence of noisy input sequences, the magnitude of degradation is different for the models of different scales. The large-scale LSRMs can maintain high stability when faced with noisy perturbations. One possible reason for this is that large-scale models are able to capture long-range dependencies among inputs, which may make them less sensitive to perturbations. By further analyzing the noisy input data, we find that the performance of the large-scale model also significantly declines when multiple perturbations occur consecutively within a sentence, which also indicates the above conjecture. This finding is useful for work on adversarial attacks and defense on recommendation models, as scaling may be a promising research direction.

\begin{figure*}[t]
\centering
\begin{minipage}{1.0\textwidth}	
        \hspace{0.2in}
	\subfigure[MovieLens-20M]{
		\label{fig:ml_trajectory}
		\includegraphics[width=0.4\textwidth]{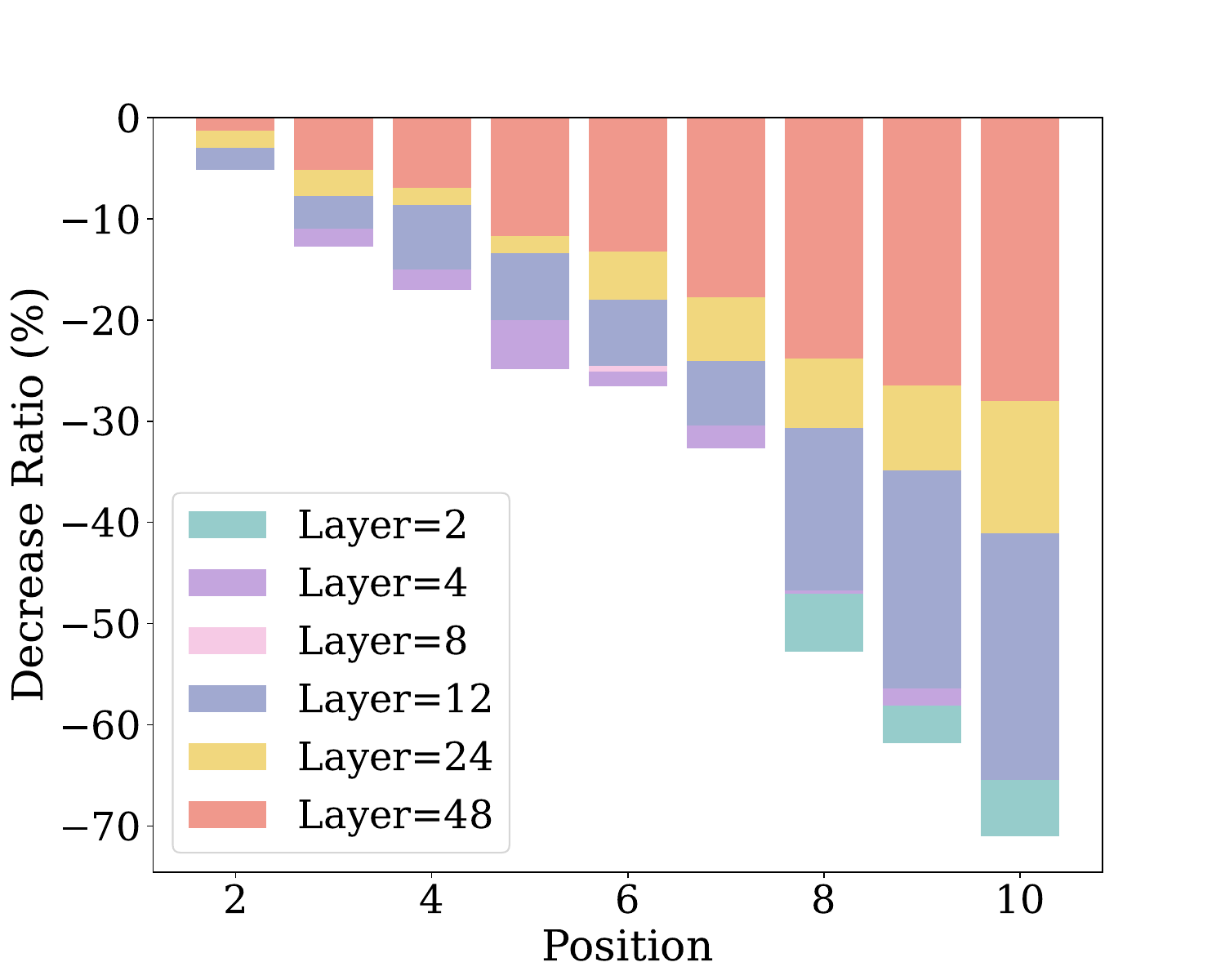}	
	}\noindent
        \hspace{0.4in}
	\subfigure[Amazon]{
		\label{fig:amazon_trajectory}
		\includegraphics[width=0.4\textwidth]{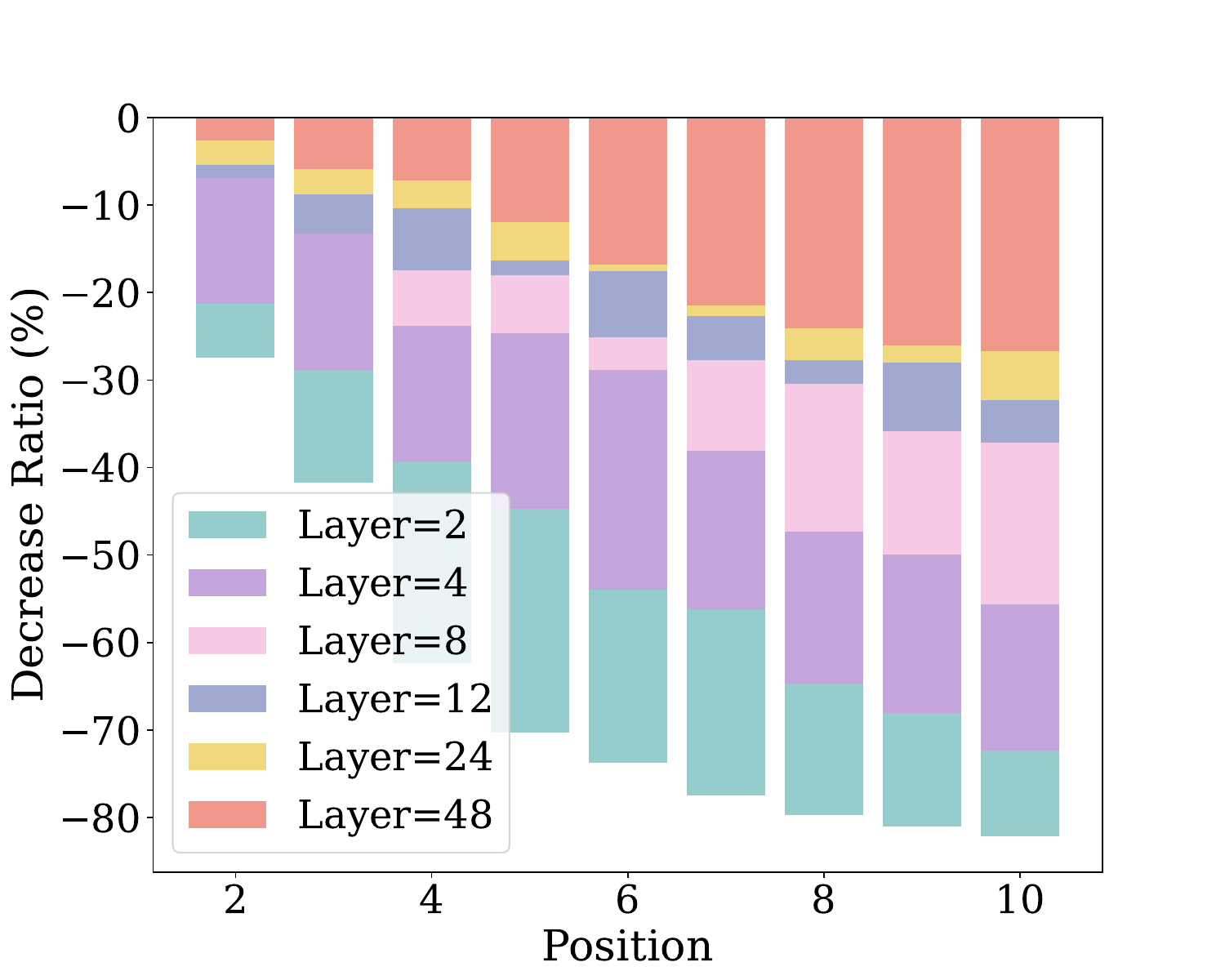}
	}
\end{minipage}
	
\caption{Performance of each model size at different positions of the user's trajectory. The $x$-axis represents the different positions of the trajectory, and the $y$-axis represents the decreased ratio of model performance~(TR metric) at each position relative to the first position.}
\label{fig:trajectory}
\end{figure*}

\subsection{User Trajectory Prediction}

In addition to the above common recommendation tasks, 
we also consider another challenging yet practical task in recommender systems, \ie user trajectory prediction (or long sequence prediction). In this task, we no longer simply predict the next item to interact with for users, but instead consider predicting a future sequence of interested items (called \emph{trajectory}) for users. This task is very meaningful in practice, since the ultimate goal of recommender systems is to establish a long-term user interest model. 


To illustrate this task, here we consider a practical example. Given a specific user $u$ and his/her interaction history $S_{h}$, our goal is to predict the future trajectory $S_{f}$ using the recommendation model. The recommendation model then generates a sequence of items based on $S_{h}$ step by step (auto-regressive), ultimately producing a predicted trajectory $S_{p}$. Note that each item in $S_{f}$ and $S_{p}$ may differ. To evaluate the long-term prediction ability of recommendation models, we can measure the similarity between $S_{f}$ and $S_{p}$. Inspired by conventional top-k metric NDCG, we use a new position-wise metric Trajectory Rank@k~(TR@k) to measure the similarity of two trajectories:

\begin{equation}
    \mathrm{TR}@k = \frac{1}{Z} \sum_{j=1}^{k} \frac{2^{I(|S_{f} \cap S_{p}[:k]|)-1}}{\log_2 (j+1)}
\end{equation}
where k denotes trajectory length, $Z = \sum_{j=1}^{k} \frac{1}{\log_2 (j+1)}$ and $I~(\cdot)$ is an indicator function.

Different from the next-item prediction task, the model predicts the next interaction each time, then adds the predicted item to the input history and further performs a new prediction based on the updated input sequence. 
Intuitively, the task becomes more difficult as the length of prediction increases, because accumulated errors are amplified when the sequence is extended. 
We vary the prediction length from 1 to 10, in which length 1 denotes the task of next-item recommendation and length 10 indicates the most difficult prediction case in our experiments.

The experimental results of different scaled LSRMs on user trajectory prediction are shown in Figure~\ref{fig:trajectory}. Firstly, we observe that as we increase the prediction length, small models experience a notable decline in performance, as they may have limited capacity to capture long-term dependencies in user trajectory data. In contrast, larger models exhibit remarkable stability in user trajectory prediction, even for long sequences. 

\vspace{1em}
To summarize, by investigating the LSRM models of different scales on five complex recommendation tasks, we find that model scalability has great potential to overcome various long-standing problems in recommendation scenarios. 

\section{Conclusion}

In this work, we investigated the scaling effect of large sequential recommendation models (LSRM) by training and testing models with parameters ranging from 98.3K to 0.8B. Overall, our empirical experiments have demonstrated that scaling laws hold in recommender systems, although with unique data characteristics, \eg data scarcity and sparsity, compared to language domains. Specially, we have conducted two groups of main experiments: 

$\bullet$ By scaling the recommendation model to a billion level, the overall performance (test loss) improves greatly (almost two times) compared to the model of traditional size. In addition, we find that it is feasible to explore predictable scaling properties by fitting scaling law curves. In the experiments, based on the performance of small models, we can accurately predict the performance of large (100$\times$ larger) models. 

$\bullet$ Furthermore, we examine the performance benefits of LSRMs in five challenging recommendation tasks, including long-tail item recommendation, cold-start user recommendation and multi-domain sequential recommendation, robustness challenges, and user long trajectory predictions. The experimental results show that large sequential models can consistently yield a stronger performance on these challenging tasks, indicating the great potential for exploring scaling effects for recommender systems. 

As for future work, we will investigate how to successfully scale up models to a larger level, and investigate more potential benefits underlying the large-scale recommendation models. To achieve this, two critical issues are in need to be studied. First, it is essential to extend the available amount of user interaction behaviors, to alleviate the data scarcity in recommender systems. Second, it is important to develop more efficient and stable optimization methods for training large recommendation models. 

\bibliographystyle{IEEEtran}
\bibliography{ref}

\begin{thebibliography}{10}
\providecommand{\url}[1]{#1}
\csname url@samestyle\endcsname
\providecommand{\newblock}{\relax}
\providecommand{\bibinfo}[2]{#2}
\providecommand{\BIBentrySTDinterwordspacing}{\spaceskip=0pt\relax}
\providecommand{\BIBentryALTinterwordstretchfactor}{4}
\providecommand{\BIBentryALTinterwordspacing}{\spaceskip=\fontdimen2\font plus
\BIBentryALTinterwordstretchfactor\fontdimen3\font minus
  \fontdimen4\font\relax}
\providecommand{\BIBforeignlanguage}[2]{{%
\expandafter\ifx\csname l@#1\endcsname\relax
\typeout{** WARNING: IEEEtran.bst: No hyphenation pattern has been}%
\typeout{** loaded for the language `#1'. Using the pattern for}%
\typeout{** the default language instead.}%
\else
\language=\csname l@#1\endcsname
\fi
#2}}
\providecommand{\BIBdecl}{\relax}
\BIBdecl

\bibitem{openai2023gpt4}
OpenAI, ``Gpt-4 technical report,'' 2023.

\bibitem{zhao2023survey}
W.~X. Zhao, K.~Zhou, J.~Li, T.~Tang, X.~Wang, Y.~Hou, Y.~Min, B.~Zhang,
  J.~Zhang, Z.~Dong \emph{et~al.}, ``A survey of large language models,''
  \emph{arXiv:2303.18223}, 2023.

\bibitem{gong2023multimodal}
T.~Gong, C.~Lyu, S.~Zhang, Y.~Wang, M.~Zheng, Q.~Zhao, K.~Liu, W.~Zhang,
  P.~Luo, and K.~Chen, ``Multimodal-gpt: A vision and language model for
  dialogue with humans,'' \emph{arXiv:2305.04790}, 2023.

\bibitem{touvron2023llama}
H.~Touvron, T.~Lavril, G.~Izacard, X.~Martinet, M.-A. Lachaux, T.~Lacroix,
  B.~Rozi{\`e}re, N.~Goyal, E.~Hambro, F.~Azhar \emph{et~al.}, ``Llama: Open
  and efficient foundation language models,'' \emph{arXiv:2302.13971}, 2023.

\bibitem{henighan2020scaling}
T.~Henighan, J.~Kaplan, M.~Katz, M.~Chen, C.~Hesse, J.~Jackson, H.~Jun, T.~B.
  Brown, P.~Dhariwal, S.~Gray \emph{et~al.}, ``Scaling laws for autoregressive
  generative modeling,'' \emph{arXiv:2010.14701}, 2020.

\bibitem{ardalani2022understanding}
N.~Ardalani, C.-J. Wu, Z.~Chen, B.~Bhushanam, and A.~Aziz, ``Understanding
  scaling laws for recommendation models,'' \emph{arXiv:2208.08489}, 2022.

\bibitem{kaplan2020scaling}
J.~Kaplan, S.~McCandlish, T.~Henighan, T.~B. Brown, B.~Chess, R.~Child,
  S.~Gray, A.~Radford, J.~Wu, and D.~Amodei, ``Scaling laws for neural language
  models,'' \emph{arXiv:2001.08361}, 2020.

\bibitem{zhai2022scaling}
X.~Zhai, A.~Kolesnikov, N.~Houlsby, and L.~Beyer, ``Scaling vision
  transformers,'' in \emph{{CVPR}}, 2022, pp. 12\,104--12\,113.

\bibitem{guo2023embedding}
X.~Guo, J.~Pan, X.~Wang, B.~Chen, J.~Jiang, and M.~Long, ``On the embedding
  collapse when scaling up recommendation models,'' \emph{arXiv:2310.04400},
  2023.

\bibitem{chitlangia2023scaling}
S.~Chitlangia, K.~R. Kesari, and R.~Agarwal, ``Scaling generative pre-training
  for user ad activity sequences,'' 2023.

\bibitem{shin2023scaling}
K.~Shin, H.~Kwak, S.~Y. Kim, M.~N. Ramstr{\"o}m, J.~Jeong, J.-W. Ha, and K.-M.
  Kim, ``Scaling law for recommendation models: Towards general-purpose user
  representations,'' in \emph{{AAAI}}, vol.~37, no.~4, 2023, pp. 4596--4604.

\bibitem{hou2022towards}
Y.~Hou, S.~Mu, W.~X. Zhao, Y.~Li, B.~Ding, and J.-R. Wen, ``Towards universal
  sequence representation learning for recommender systems,'' in \emph{KDD},
  2022, pp. 585--593.

\bibitem{li2023text}
J.~Li, M.~Wang, J.~Li, J.~Fu, X.~Shen, J.~Shang, and J.~McAuley, ``Text is all
  you need: Learning language representations for sequential recommendation,''
  \emph{arXiv:2305.13731}, 2023.

\bibitem{hou2023learning}
Y.~Hou, Z.~He, J.~McAuley, and W.~X. Zhao, ``Learning vector-quantized item
  representation for transferable sequential recommenders,'' in \emph{WWW},
  2023, pp. 1162--1171.

\bibitem{muennighoff2023scaling}
N.~Muennighoff, A.~M. Rush, B.~Barak, T.~L. Scao, A.~Piktus, N.~Tazi,
  S.~Pyysalo, T.~Wolf, and C.~Raffel, ``Scaling data-constrained language
  models,'' \emph{arXiv:2305.16264}, 2023.

\bibitem{liu2023dropout}
Z.~Liu, Z.~Xu, J.~Jin, Z.~Shen, and T.~Darrell, ``Dropout reduces
  underfitting,'' \emph{arXiv:2303.01500}, 2023.

\bibitem{he2020lightgcn}
X.~He, K.~Deng, X.~Wang, Y.~Li, Y.~Zhang, and M.~Wang, ``Lightgcn: Simplifying
  and powering graph convolution network for recommendation,'' in
  \emph{{SIGIR}}, 2020, pp. 639--648.

\bibitem{kang2018self}
W.-C. Kang and J.~McAuley, ``Self-attentive sequential recommendation,'' in
  \emph{{ICDM}}.

\bibitem{sun2019bert4rec}
F.~Sun, J.~Liu, J.~Wu, C.~Pei, X.~Lin, W.~Ou, and P.~Jiang, ``Bert4rec:
  Sequential recommendation with bidirectional encoder representations from
  transformer,'' in \emph{{CIKM}}, 2019, pp. 1441--1450.

\bibitem{lin2022improving}
Z.~Lin, C.~Tian, Y.~Hou, and W.~X. Zhao, ``Improving graph collaborative
  filtering with neighborhood-enriched contrastive learning,'' in \emph{{WWW}},
  2022, pp. 2320--2329.

\bibitem{rendle2010factorizing}
S.~Rendle, C.~Freudenthaler, and L.~Schmidt-Thieme, ``Factorizing personalized
  markov chains for next-basket recommendation,'' in \emph{{WWW}}, 2010, pp.
  811--820.

\bibitem{hidasi2015session}
B.~Hidasi, A.~Karatzoglou, L.~Baltrunas, and D.~Tikk, ``Session-based
  recommendations with recurrent neural networks,'' \emph{arXiv:1511.06939},
  2015.

\bibitem{tang2018personalized}
J.~Tang and K.~Wang, ``Personalized top-n sequential recommendation via
  convolutional sequence embedding,'' in \emph{{WSDM}}, 2018, pp. 565--573.

\bibitem{wu2019session}
S.~Wu, Y.~Tang, Y.~Zhu, L.~Wang, X.~Xie, and T.~Tan, ``Session-based
  recommendation with graph neural networks,'' in \emph{{AAAI}}, vol.~33,
  no.~01, 2019, pp. 346--353.

\bibitem{chang2021sequential}
J.~Chang, C.~Gao, Y.~Zheng, Y.~Hui, Y.~Niu, Y.~Song, D.~Jin, and Y.~Li,
  ``Sequential recommendation with graph neural networks,'' in \emph{{SIGIR}},
  2021, pp. 378--387.

\bibitem{zhou2022filter}
K.~Zhou, H.~Yu, W.~X. Zhao, and J.-R. Wen, ``Filter-enhanced mlp is all you
  need for sequential recommendation,'' in \emph{{WWW}}, 2022, pp. 2388--2399.

\bibitem{vaswani2017attention}
A.~Vaswani, N.~Shazeer, N.~Parmar, J.~Uszkoreit, L.~Jones, A.~N. Gomez,
  {\L}.~Kaiser, and I.~Polosukhin, ``Attention is all you need,''
  \emph{{NeurIPS}}, vol.~30, 2017.

\bibitem{li2021lightweight}
Y.~Li, T.~Chen, P.-F. Zhang, and H.~Yin, ``Lightweight self-attentive
  sequential recommendation,'' in \emph{{CIKM}}, 2021, pp. 967--977.

\bibitem{hou2022core}
Y.~Hou, B.~Hu, Z.~Zhang, and W.~X. Zhao, ``Core: Simple and effective
  session-based recommendation within consistent representation space,'' in
  \emph{{SIGIR}}, 2022.

\bibitem{fan2021lighter}
X.~Fan, Z.~Liu, J.~Lian, W.~X. Zhao, X.~Xie, and J.-R. Wen, ``Lighter and
  better: low-rank decomposed self-attention networks for next-item
  recommendation,'' in \emph{{SIGIR}}, 2021, pp. 1733--1737.

\bibitem{mhaskar1996neural}
H.~N. Mhaskar, ``Neural networks for optimal approximation of smooth and
  analytic functions,'' \emph{Neural computation}, vol.~8, no.~1, pp. 164--177,
  1996.

\bibitem{brown2020language}
T.~Brown, B.~Mann, N.~Ryder, M.~Subbiah, J.~D. Kaplan, P.~Dhariwal,
  A.~Neelakantan, P.~Shyam, G.~Sastry, A.~Askell \emph{et~al.}, ``Language
  models are few-shot learners,'' \emph{{NeurIPS}}, vol.~33, pp. 1877--1901,
  2020.

\bibitem{tang2023does}
R.~Tang, X.~Han, X.~Jiang, and X.~Hu, ``Does synthetic data generation of llms
  help clinical text mining?'' \emph{arXiv:2303.04360}, 2023.

\bibitem{nov2023putting}
O.~Nov, N.~Singh, and D.~M. Mann, ``Putting chatgpt's medical advice to the
  (turing) test,'' \emph{medRxiv}, pp. 2023--01, 2023.

\bibitem{malinka2023educational}
K.~Malinka, M.~Peres{\'\i}ni, A.~Firc, O.~Hujnak, and F.~Janus, ``On the
  educational impact of chatgpt: Is artificial intelligence ready to obtain a
  university degree?'' in \emph{ITiCSE}, 2023, pp. 47--53.

\bibitem{yang2023fingpt}
H.~Yang, X.-Y. Liu, and C.~D. Wang, ``Fingpt: Open-source financial large
  language models,'' \emph{arXiv:2306.06031}, 2023.

\bibitem{sun2023short}
Z.~Sun, ``A short survey of viewing large language models in legal aspect,''
  \emph{arXiv:2303.09136}, 2023.

\bibitem{zhang2023one}
C.~Zhang, C.~Zhang, C.~Li, Y.~Qiao, S.~Zheng, S.~K. Dam, M.~Zhang, J.~U. Kim,
  S.~T. Kim, J.~Choi \emph{et~al.}, ``One small step for generative ai, one
  giant leap for agi: A complete survey on chatgpt in aigc era,''
  \emph{arXiv:2304.06488}, 2023.

\bibitem{wu2023survey}
L.~Wu, Z.~Zheng, Z.~Qiu, H.~Wang, H.~Gu, T.~Shen, C.~Qin, C.~Zhu, H.~Zhu,
  Q.~Liu \emph{et~al.}, ``A survey on large language models for
  recommendation,'' \emph{arXiv:2305.19860}, 2023.

\bibitem{lin2023can}
J.~Lin, X.~Dai, Y.~Xi, W.~Liu, B.~Chen, X.~Li, C.~Zhu, H.~Guo, Y.~Yu, R.~Tang
  \emph{et~al.}, ``How can recommender systems benefit from large language
  models: A survey,'' \emph{arXiv:2306.05817}, 2023.

\bibitem{fan2023recommender}
W.~Fan, Z.~Zhao, J.~Li, Y.~Liu, X.~Mei, Y.~Wang, J.~Tang, and Q.~Li,
  ``Recommender systems in the era of large language models (llms),''
  \emph{arXiv:2307.02046}, 2023.

\bibitem{li2023large}
L.~Li, Y.~Zhang, D.~Liu, and L.~Chen, ``Large language models for generative
  recommendation: A survey and visionary discussions,''
  \emph{arXiv:2309.01157}, 2023.

\bibitem{hou2023large}
Y.~Hou, J.~Zhang, Z.~Lin, H.~Lu, R.~Xie, J.~McAuley, and W.~X. Zhao, ``Large
  language models are zero-shot rankers for recommender systems,''
  \emph{arXiv:2305.08845}, 2023.

\bibitem{zhang2023recommendation}
J.~Zhang, R.~Xie, Y.~Hou, W.~X. Zhao, L.~Lin, and J.-R. Wen, ``Recommendation
  as instruction following: A large language model empowered recommendation
  approach,'' \emph{arXiv:2305.07001}, 2023.

\bibitem{wang2023zero}
L.~Wang and E.-P. Lim, ``Zero-shot next-item recommendation using large
  pretrained language models,'' \emph{arXiv:2304.03153}, 2023.

\bibitem{liu2023chatgpt}
J.~Liu, C.~Liu, R.~Lv, K.~Zhou, and Y.~Zhang, ``Is chatgpt a good recommender?
  a preliminary study,'' \emph{arXiv:2304.10149}, 2023.

\bibitem{harte2023leveraging}
J.~Harte, W.~Zorgdrager, P.~Louridas, A.~Katsifodimos, D.~Jannach, and
  M.~Fragkoulis, ``Leveraging large language models for sequential
  recommendation,'' in \emph{RecSys}, 2023, pp. 1096--1102.

\bibitem{bao2023tallrec}
K.~Bao, J.~Zhang, Y.~Zhang, W.~Wang, F.~Feng, and X.~He, ``Tallrec: An
  effective and efficient tuning framework to align large language model with
  recommendation,'' \emph{arXiv:2305.00447}, 2023.

\bibitem{zhang2023agentcf}
J.~Zhang, Y.~Hou, R.~Xie, W.~Sun, J.~McAuley, W.~X. Zhao, L.~Lin, and J.-R.
  Wen, ``Agentcf: Collaborative learning with autonomous language agents for
  recommender systems,'' \emph{arXiv:2310.09233}, 2023.

\bibitem{wolf-etal-2020-transformers}
\BIBentryALTinterwordspacing
T.~Wolf, L.~Debut, V.~Sanh, J.~Chaumond, C.~Delangue, A.~Moi, P.~Cistac,
  T.~Rault, R.~Louf, M.~Funtowicz, J.~Davison, S.~Shleifer, P.~von Platen,
  C.~Ma, Y.~Jernite, J.~Plu, C.~Xu, T.~L. Scao, S.~Gugger, M.~Drame, Q.~Lhoest,
  and A.~M. Rush, ``Transformers: State-of-the-art natural language
  processing,'' in \emph{{EMNLP}}.\hskip 1em plus 0.5em minus 0.4em\relax
  Online: Association for Computational Linguistics, Oct. 2020, pp. 38--45.
  [Online]. Available:
  \url{https://www.aclweb.org/anthology/2020.emnlp-demos.6}
\BIBentrySTDinterwordspacing

\bibitem{zhao2021recbole}
W.~X. Zhao, S.~Mu, Y.~Hou, Z.~Lin, Y.~Chen, X.~Pan, K.~Li, Y.~Lu, H.~Wang,
  C.~Tian \emph{et~al.}, ``Recbole: Towards a unified, comprehensive and
  efficient framework for recommendation algorithms,'' in \emph{{CIKM}}, 2021,
  pp. 4653--4664.

\bibitem{he2016deep}
K.~He, X.~Zhang, S.~Ren, and J.~Sun, ``Deep residual learning for image
  recognition,'' in \emph{Proceedings of the IEEE conference on computer vision
  and pattern recognition}, 2016, pp. 770--778.

\bibitem{baevski2018adaptive}
A.~Baevski and M.~Auli, ``Adaptive input representations for neural language
  modeling,'' \emph{arXiv:1809.10853}, 2018.

\bibitem{srivastava2014dropout}
N.~Srivastava, G.~Hinton, A.~Krizhevsky, I.~Sutskever, and R.~Salakhutdinov,
  ``Dropout: a simple way to prevent neural networks from overfitting,''
  \emph{The journal of machine learning research}, vol.~15, no.~1, pp.
  1929--1958, 2014.

\bibitem{hoffmann2022training}
J.~Hoffmann, S.~Borgeaud, A.~Mensch, E.~Buchatskaya, T.~Cai, E.~Rutherford,
  D.~d.~L. Casas, L.~A. Hendricks, J.~Welbl, A.~Clark \emph{et~al.}, ``Training
  compute-optimal large language models,'' \emph{arXiv:2203.15556}, 2022.

\bibitem{kingma2014adam}
D.~P. Kingma and J.~Ba, ``Adam: A method for stochastic optimization,''
  \emph{arXiv:1412.6980}, 2014.

\bibitem{robbins1951stochastic}
H.~Robbins and S.~Monro, ``A stochastic approximation method,'' \emph{The
  annals of mathematical statistics}, pp. 400--407, 1951.

\bibitem{keskar2017improving}
N.~S. Keskar and R.~Socher, ``Improving generalization performance by switching
  from adam to sgd,'' \emph{arXiv preprint arXiv:1712.07628}, 2017.

\bibitem{harper2015movielens}
F.~M. Harper and J.~A. Konstan, ``The movielens datasets: History and
  context,'' \emph{{TiiS}}, vol.~5, no.~4, pp. 1--19, 2015.

\bibitem{ni2019justifying}
J.~Ni, J.~Li, and J.~McAuley, ``Justifying recommendations using
  distantly-labeled reviews and fine-grained aspects,'' in
  \emph{{EMNLP-IJCNLP}}, 2019, pp. 188--197.

\bibitem{yu2020semi}
W.~Yu, X.~Lin, J.~Ge, W.~Ou, and Z.~Qin, ``Semi-supervised collaborative
  filtering by text-enhanced domain adaptation,'' in \emph{KDD}, 2020, pp.
  2136--2144.

\bibitem{gage1994new}
P.~Gage, ``A new algorithm for data compression,'' \emph{C Users Journal},
  vol.~12, no.~2, pp. 23--38, 1994.

\bibitem{carlini2022quantifying}
N.~Carlini, D.~Ippolito, M.~Jagielski, K.~Lee, F.~Tramer, and C.~Zhang,
  ``Quantifying memorization across neural language models,''
  \emph{arXiv:2202.07646}, 2022.

\bibitem{tirumala2022memorization}
K.~Tirumala, A.~Markosyan, L.~Zettlemoyer, and A.~Aghajanyan, ``Memorization
  without overfitting: Analyzing the training dynamics of large language
  models,'' \emph{NeurIPS}, vol.~35, pp. 38\,274--38\,290, 2022.

\bibitem{cao2022contrastive}
J.~Cao, X.~Cong, J.~Sheng, T.~Liu, and B.~Wang, ``Contrastive cross-domain
  sequential recommendation,'' in \emph{CIKM}, 2022, pp. 138--147.

\bibitem{tang2023one}
Z.~Tang, Z.~Huan, Z.~Li, X.~Zhang, J.~Hu, C.~Fu, J.~Zhou, and C.~Li, ``One
  model for all: Large language models are domain-agnostic recommendation
  systems,'' \emph{arXiv:2310.14304}, 2023.

\bibitem{tan2023towards}
J.~Tan, S.~Heinecke, Z.~Liu, Y.~Chen, Y.~Zhang, and H.~Wang, ``Towards more
  robust and accurate sequential recommendation with cascade-guided adversarial
  training,'' \emph{arXiv:2304.05492}, 2023.

\bibitem{he2018adversarial}
X.~He, Z.~He, X.~Du, and T.-S. Chua, ``Adversarial personalized ranking for
  recommendation,'' in \emph{The 41st International ACM SIGIR conference on
  research \& development in information retrieval}, 2018, pp. 355--364.

\bibitem{o2004collaborative}
M.~O'Mahony, N.~Hurley, N.~Kushmerick, and G.~Silvestre, ``Collaborative
  recommendation: A robustness analysis,'' \emph{TOIT}, vol.~4, no.~4, pp.
  344--377, 2004.

\bibitem{o2005trust}
J.~O'Donovan and B.~Smyth, ``Trust in recommender systems,'' in \emph{IUI},
  2005, pp. 167--174.

\end{thebibliography}

\end{document}